\documentclass{aa}
\usepackage{graphicx}
\usepackage{natbib}

\begin{document}
   \title{A compact dust shell in the symbiotic system \object{HM\,Sge}\thanks{Based on observations made with the Very Large Telescope Interferometer at Paranal Observatory under programs 075.D-0484 and 077.D-0216}}

\titlerunning{The dust shell of \object{HM\,Sge}}

\authorrunning{Sacuto, S. et al.} 

   \author{S.~Sacuto\inst{1}, O.~Chesneau\inst{1}, M.~Vannier\inst{2}, P.~Cruzal\`{e}bes\inst{1}}

   \offprints{S.~Sacuto}

\institute{Observatoire de la C\^{o}te d'Azur, Dpt. Gemini-CNRS-UMR 6203, Avenue Copernic, F-06130 Grasse\\
\email{stephane.sacuto@obs-azur.fr}
\and
ESO, Alonso de Cordova 3107, Vitacura. Casilla 19001, Santiago 19, Chile.}

   \date{Received; accepted }

  \abstract
{}
{We present high spatial resolution observations of the mid-infrared core of the dusty symbiotic system \object{HM\,Sge}\thanks{Reduced visibilities and differential phases are available in electronic form at the CDS via anonymous ftp to cdsarc.u-strasbg.fr (130.79.128.5) or via http://cdsweb.u-strasbg.fr/cgi-bin/qcat?J/A+A/}.}
{The MIDI interferometer was used with the VLT Unit Telescopes and Auxiliary Telescopes providing baselines oriented from PA=42$^\circ$ to 105$^\circ$. The MIDI visibilities are compared with the ones predicted in the frame of various spherical dust shells published in the literature involving single or double dusty shells intended to account for the influence of the hot White Dwarf.}
{The mid-IR environment is unresolved by a 8m telescope (resolution$\sim$300\,mas) and the MIDI spectrum exhibits a level similar to the ISO spectra recorded 10\,yr ago. The estimated Gaussian Half Width at Half Maximum of the shell of 7.8$\pm$1.3\,mas (12AU, assuming a distance of 1.5kpc) in the 8-9$\mu$m range, and 11.9$\pm$1.3\,mas (18AU) in the 11-12$\mu$m range, are much smaller than the angular separation between the Mira and the White Dwarf of 40\,mas (60AU). The discrepancies between the HWHM at different angle orientations suggest an increasing level of asymmetry from 13 to 8$\mu$m. The observations are surprisingly well fitted by the densest (optically thick in the N band) and smallest spherical model published in the literature based on the ISO data, although such a model does not account for the variations of near-IR photometry due to the Mira pulsation cycle suggesting a much smaller optical thickness. These observations also discard the two shells models, developed in an attempt to take into account the effect of the White Dwarf illumination onto the dusty wind of the Mira. These models are too extended, and lead to a level of asymmetry of the dusty environment tightly constrained by the MIDI visibilities. These observations show that a high rate of dust formation is occurring in the vicinity of the Mira which seems to be not highly perturbed by the hot companion.}
{}


   \maketitle
%


\section{Introduction}
\object{HM\,Sge} is a D-type (dust forming) symbiotic system that erupted as a symbiotic nova in 1975 \citep{doku76}, evolving from a 17th to a 11th magnitude star with a rich emission line spectrum. The cool component is a Mira star and the hot component is a White Dwarf (WD) which had escaped detection enshrouded in the dense Mira envelop before this dramatic event. 

\object{HM\,Sge} was observed in IR from the ground and also with the IRAS and ISO satellites. The observations show a spectrum dominated by silicate dust whose flux evolved deeply in about 30\,yrs. Many papers investigated the time variable dusty environment and tried to evaluate the impact of the temperature and luminosity evolution of the WD on the dust \citep{doku76,wenz76,ciat77,davi78,ciat78,kwok79,brya91,yudi94,bogd01,schi01}. 
The strong infrared excess detected shortly after the outburst indicates the presence of a rather dense dust envelope. The dust content of the system was severely affected by the nova outburst and it was soon observed that a process of dissipation of the dust was on the way \citep{tara82,tara83}, despite some obscuration events \citep{muna89}. The IRAS spectrum obtained in 1983 is about twice the ISO spectra obtained in 1996-1997 and this flux difference
was attributed to a long-term variation effect and not to a beam size variation. \citet{bogd01} attributed these changes to an increase of the concentration and decrease of the temperature of the dust close to the Mira star, associated with the temperature increase and luminosity decrease of the White Dwarf. In the other hand, \citet{schm00} find that the red giant spectrum has strongly brightened and find that the dust column in the line of sight has markedly
decreased. It is difficult to assess whether this is a long-term evolution or an other event related to the presumably patchy circumstellar environment. This last hypothesis is favored by \citet{schm00} and it must be noticed that longer wavelength emission is less sensitive to the clumpiness of the environment which probably implies that the conclusions of \citet{bogd01} and \citet{schm00} are not necessarily incompatible.

Classically, most of the studies evaluated the infrared excess by fitting model spectral energy distributions (SEDs) to the instantaneous observed ones. The common model is a single dust shell, whose composition is dominated by silicates (but some studies involve a contribution from carbon \citep{brya91}). However, this simple geometry could be considered as a crude approximation of a putatively complex geometry and dust composition of the mid-IR emitting regions. As an attempt to overcome this problem, more complex models involving two shells were used: a cool one surrounding the Mira star and a second, more extended that sees mostly the WD flux \citep{bogd01,schi01}. The best argument for a two shell model is that the dust-rich area emission is affected by the
combined effects of a ionizing UV irradiation from the hot
primary on the expanding dust shell of the red giant, and a
compression wave of the colliding winds in the transition zone. The conclusions of such approaches appear in contradiction to each other: whereas \citet{schi01} found a better match of the data with a two-component model of dust
distribution, \citet{bogd01} are lead to the opposite conclusion and claim that current models with only a single Mira-type source are preferable. It is important to notice that both studies use the same ISO spectra as constraints and that they are able to provide satisfactory fits of the SED with very different model parameters. It is obvious that the single constraint of a SED is not able to fully constrain a putatively complex dust environment whose composition, geometry and even the luminosity of the heating sources is poorly known. An additional difficulty is that the mid-IR observations are to date scarce and secured only with relatively extended beams. Moreover, \object{HM\,Sge} is not in an equilibrium state (Mira pulsations, wind interactions,
dissipation/formation of obscuring dust,\dots) and this picture might well have evolved over the past few years since the ISO observations in 1996-1997. There is also a need for a SED taken with a much smaller beam than the ISO one which would allow to isolate the signature of the most recently formed dust.

We present in this article the first mid-infrared interferometric observations of this system taken with the MIDI/VLTI instrument recorded in 2005/2006 providing a spectrum update from a 8m telescope beam (FWHM$\sim$300\,mas) and a 1.5m telescope beam (FWHM$\sim$1.5$\arcsec$), and measurements of spatially correlated flux in interferometric mode that provide high spatial resolution ($\sim$20\,mas) and multi-spectral ($\lambda / \Delta \lambda \sim 30$) information on the mid-IR source. The flux of VLT single dish spectrum is directly comparable to the the ISO ones (see Fig.~\ref{fig:spectra}). This is a clear indication that, first, most of the ISO emission observed 8yr before probably originated from a spatially compact region and second, that the dust content of the system seems not to have changed significantly. As a consequence, we assume in the following study that the spatial distributions expected from the different models developed by \citet{schi01} and \citet{bogd01} can be checked against these high spatial resolution observations and that the MIDI spectrum can be complemented with the ISO spectra to construct the current SED of the system at longer infrared wavelengths. The outline of the paper is the following: we present first the MIDI observations in Sec.\ref{sect:obs} and we summarize the published parameters of the models developed by \citet{schi01} and \citet{bogd01} in Sec.\ref{sect:SEDmodels}. Then in Sec.\ref{sect:Vismodels}, the expected visibilities from the models are compared to the MIDI observations (the detail of the calculations is shown in appendix). We propose some improvements for the best model in Sec.\ref{sect:modelImp}. We discuss the limitations of the approach based on spherically symmetric dust shells for the study of a spatially complex source in Sec.\ref{sect:discuss}. Finally we summarize and conclude in Sec.\ref{sect:conclusion}.

\section{Observations}
\label{sect:obs}
The Very Large Telescope Interferometer (VLTI) of ESO's Paranal Observatory has been used with MIDI, the MID-infrared Interferometric recombiner \citep{lein03}. MIDI combines the light of two telescopes and provides spectrally resolved visibilities in the N band atmospheric window. 

The observations of \object{HM\,Sge} were conducted with the VLT Unit Telescopes (UTs) UT2, UT3 and UT4, providing projected baselines in the range of 32-59 meters and the Auxiliary Telescopes (ATs) E0 and G0, providing projected baselines in the range of 13.5-16 meters.

The UTs observations were made during the nights of July 23-24th 2005, May 17th 2006 and June 11th 2006. The phase of the Mira during the UTs MIDI observations was 0.75 in 2005 and between 0.31-0.36 in 2006 \citep{yudi94}.
The data were recorded with different projected baselines, (46.5m, 44$^\circ$), (32.1m, 42$^\circ$), (37.8m, 47$^\circ$), (46.2m, 47$^\circ$), (59.3, 105$^\circ$) and (46.8, 101$^\circ$). 
The ATs observations were made during the nights of May 21th 2006, May 27th 2006 and June 16th 2006. The phase of the Mira during the ATs MIDI observations was between 0.32-0.37 in 2006 \citep{yudi94}.
The data were recorded with different projected baselines, (13.5m, 80$^\circ$), (15.8m, 69$^\circ$), (15.8m, 75$^\circ$) and (15.3m, 77$^\circ$) but the night of May 21th 2006 was discarded due to the misalignment of the star (see below).
All the observations were made under good atmospheric conditions ($\overline{seeing}$$\sim$1.01$\arcsec$) with the worst seeing ($\sim$1.75$\arcsec$) during the night of July 23th 2005\footnote{The Adaptive Optics device MACAO dedicated to the VLTI provide diffraction limited images in N band for the UTs even for the worst condition encountered on this observation}.

The data reduction softwares\footnote{http://www.mpia-hd.mpg.de/MIDISOFT/, http://www.strw.leidenuniv.nl/$\sim$nevec/MIDI/} MIA and EWS \citep{jaff04} were used to reduce the spectra and visibilities \citep{ches05a}. 
Chopped acquisition images are recorded (f=2Hz, 2000 frames, 4~ms per frame, 98\,mas per pixel) for the fine acquisition of the target. The acquisition filter was a N band filter. The prism of MIDI was used providing a spectral dispersion of about 30.

Tab.~\ref{tab:journal2} presents the journal of the interferometric observations. The calibrators, \object{HD188512} (G8IV, diam=1.98$\pm$0.02\,mas), \object{HD187642} (A7V, diam=3.22$\pm$0.01\,mas) and \object{HD206778} (K2Ib, diam=8.38$\pm$0.09\,mas) were observed right before or after each science target observations. 

\label{sec:acqim}

\begin{table}[h]
\caption{\label{tab:journal2}Journal of observations: MIDI/UT2-UT3/UT3-UT4/E0-G0. The calibrators used to calibrate the visibilities are given below the science target. The phase of the Mira ($\varphi_{ Mira}$) during the observations is indicated. The length and position angle of the projected baseline are also indicated.}
\vspace{0.3cm}
\begin{center}
\begin{tabular}{ccccc}
\hline
\hline
{\tiny Star} & {\tiny UT date \& Time} & {\tiny $\varphi_{Mira}$} & {\tiny Base[m]} & {\tiny PA [deg]} \\
\hline
{\tiny \object{HM\,Sge}} & {\tiny 2005-07-23 06:48:42} & 0.75 & 46.5 & 44 \\
{\tiny \object{HD188512}} & {\tiny 2005-07-23 07:12:10} & - & - & - \\
\hline
{\tiny \object{HM\,Sge}} & {\tiny 2005-07-24 02:41:11} & 0.75 & 32.1 & 42 \\
{\tiny \object{HD188512}} & {\tiny 2005-07-24 02:58:55} & - & - & - \\
\hline
{\tiny \object{HM\,Sge}} & {\tiny 2005-07-24 03:42:18} & 0.75 & 37.8 & 47 \\
{\tiny \object{HD188512}} & {\tiny 2005-07-24 04:03:14} & - & - & - \\
\hline
{\tiny \object{HM\,Sge}} & {\tiny 2005-07-24 06:14:49} & 0.75 & 46.2 & 47 \\
{\tiny \object{HD188512}} & {\tiny 2005-07-24 06:37:58} & - & - & - \\
\\\hline\hline\\
\hline
{\tiny \object{HM\,Sge}} & {\tiny 2006-05-17 09:04:02} & 0.31 & 59.3 & 105 \\
{\tiny \object{HD187642}} & {\tiny 2006-05-17 09:27:14} & - & - & - \\
\hline
{\tiny \object{HM\,Sge}} & {\tiny 2006-06-11 09:05:09} & 0.36 & 46.8 & 101 \\
{\tiny \object{HD187642}} & {\tiny 2006-06-11 08:42:52} & - & - & - \\
\hline
{\tiny \object{HM\,Sge}} & {\tiny 2006-05-27 08:17:39} & 0.33 & 15.8 & 75 \\
{\tiny \object{HD206778}} & {\tiny 2006-05-27 08:39:57} & - & - & - \\
\hline
{\tiny \object{HM\,Sge}} & {\tiny 2006-06-16 06:28:01} & 0.37 & 15.3 & 77 \\
{\tiny \object{HD206778}} & {\tiny 2006-06-16 05:53:35} & - & - & - \\
\hline
\end{tabular}
\end{center}
\end{table}

\normalsize

The spectral template from \citet{cohe99} of \object{HD180711} (G9III) was scaled to the 12$\mu$m flux of the interferometric calibrator \object{HD188512} for absolute flux calibration. In interferometry, in parallel with measurements of the source, measurements of the nearby calibrator are taken. Because the difference in airmass did not exceed 0.1, no airmass correction was applied to the corrected spectra.

\begin{figure}
\begin{center}
\includegraphics[width=8.cm]{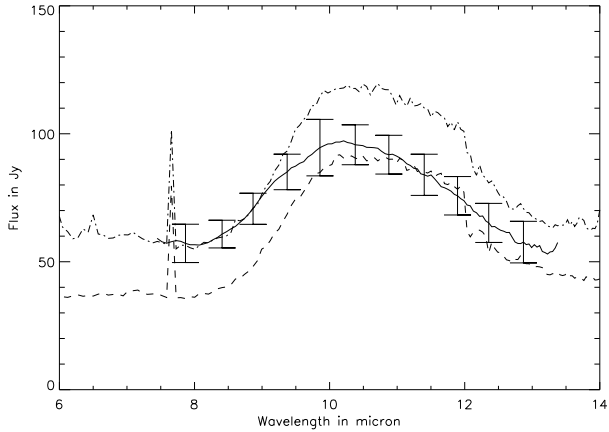}
\end{center}
\caption{\label{fig:spectra} The solid line with error bars is the MIDI flux (Mira pulsation phase, $\varphi$=0.75) and the dashed and dashed-dotted lines are the ISO/SWS spectra (upper one: 1996, $\varphi$=0.65, bottom one: 1997, $\varphi$=1.08)}

\end{figure}

\begin{figure}
\begin{center}
\includegraphics[width=9.cm]{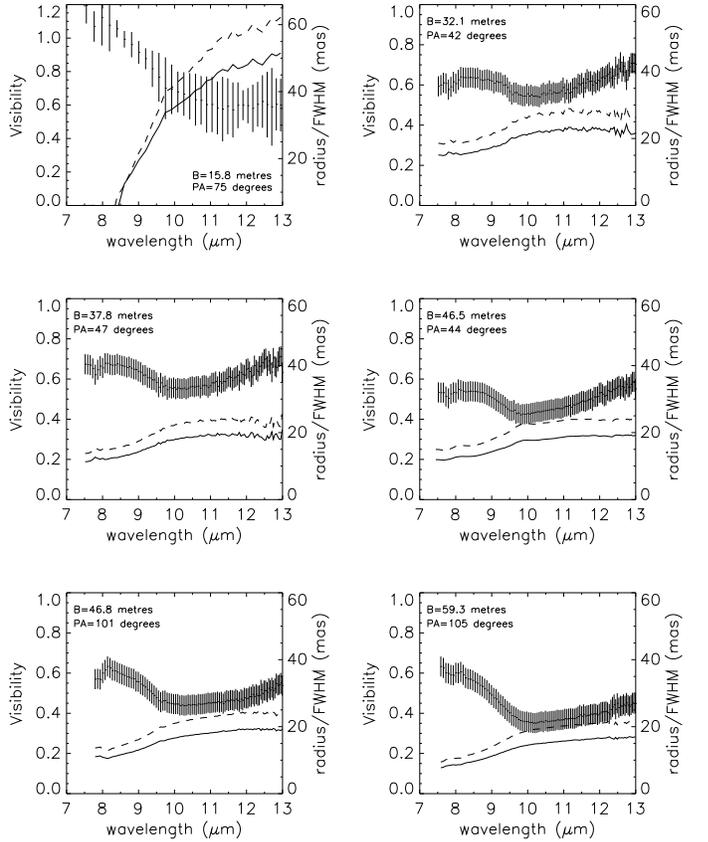}
\end{center}
\caption{\label{fig:DU_Gauss} MIDI visibilities (error bars) for the six projected baselines. The solid lines correspond to the uniform disk angular radius in mas and the dashed lines correspond to the FWHM in mas of the Gaussian distribution (both to be read from the scale on the right axis), computed from $V$ at each wavelength.}

\end{figure}

The MIDI spectrum compared to the ISO spectra taken at two different phases of the Mira is shown in Fig.~\ref{fig:spectra}. The level of flux received by MIDI is equivalent to the flux level of ISO spectrum, implying that most of the N flux is concentrated within the 300\,mas beam of a 8m telescope and also that the level of emission has probably not significantly changed since the ISO observations (1996-1997). We notice that the flux error bars of the photometry from UTs are below 8\% level. The shape of the silicate feature in the MIDI spectrum is different from that of the ISO spectra: the flux level at 8$\mu$m is close to the upper ISO spectrum whereas the flux longward 12$\mu$m is closer to the lower ISO spectrum. It is difficult to state whether this is due to the signature of the dusty source at this particular pulsation phase or the consequence of a longer term evolution. 

We used all calibrators of each observing night to evaluate the quality of the calibration. The data were reduced with both MIA and EWS. For the UTs data, the quality is uniformly good and we decided to
apply a uniform multiplicative error of 10\% as a conservative estimate. The data reduction of the ATs data was much more difficult for several reasons: the alignment of the source was never optimal and strong biases in the extracted photometries were observed, affecting more the edges of the N band (7.5-9$\mu$m and 12-13.5$\mu$m). Moreover, the data reduction softwares are not currently optimized for these telescopes and developments are under way.
As a consequence, we decided to show these data, but put only little weight on them compared to the more robust UTs data in the interpretation process.
Some MIDI visibility curves were merged and averaged.
We consider the two UTs observations of the July 23th 2005 at 06:48 and July 24th 2005 at 06:14 as approximatively redundant because of their small baseline and position angle (PA) variation ($<$1\% and $<$7\%, respectively). This is also the case for the 2 ATs observations (May 27th 2006 and June 16th 2006) which have been averaged due to their small projected baselines and also their bad quality. The data from May 21th 2006 were disregarded due to alignment problems. The other ATs data were reduced, calibrated and merged in a single curve in Fig.~\ref{fig:DU_Gauss}. We notice that the error bars at the edges of the N band, strongly affected by the photometric biaises are probably underevaluated. The six remaining calibrated visibility data sets of \object{HM\,Sge} over the wavelength range from 7.5 to 13$\mu$m are shown in Fig.~\ref{fig:DU_Gauss}. The figure also shows the plot of the equivalent uniform-disk angular radius and FWHM of the Gaussian distribution calculated from the visibilities at each spectral channel. The size of both models increases from 7.5 to 10$\mu$m and is approximately constant between 10 and 13$\mu$m.
Notice that the wavelength range has been slightly increased to show the MIDI visibilities between 7 and 8$\mu$m, at the edge of the atmospheric window. The quality of the data is good, and we can see a decrease of visibility that corresponds to the very strong emission of the forbidden line [NeVI] at 7.65$\mu$m (see Fig.~\ref{fig:spectra}). This narrow line is observed in the best quality MIDI spectra but almost disappeared in the mean spectrum presented in Fig.~\ref{fig:spectra} due to small mismatch of the source position on the slit. This decrease can be interpreted as a much larger extension for the line emitting region. This is not investigated further in the paper due to the limited spectral resolution and the potentially large errors at this wavelength. Finally, all visibility and differential phase data as well as all the characteristics of the observations are available from the CDS (Centre de Donn\'{e}es astronomiques de Strasbourg); all data products are stored in the FITS-based, optical interferometry data exchange format (OI-FITS), described in \citet{paul05}.

\section{Spectrophotometric models}
\label{sect:SEDmodels}

The infrared observations (mostly from IRAS and ISO) were interpreted within the frame of spherical models generated by the DUSTY package \citep{ivez99}. We first reproduce the SEDs of models published by \citet{schi01} (S01) and \citet{bogd01} (B01). Then, still with the DUSTY code, we generate the corresponding synthetic spectrally-dispersed visibility profiles throughout the N band (7.5-13$\mu$m) and compare them with the MIDI visibilities for each baselines. The aims of our study are threefold: first, finding the model that provides the closest match with the data; second improving, if possible, the model parameters; and third, based on the best model, investigating a putative departure of spherical symmetry of the source.

\subsection{Description of the models}

The models generated by S01 and B01 were based on the fit of the ISO SWS spectra of \object{HM\,Sge} obtained in 10/01/1996 and 05/16/1997 \citep{graa96,sala97}. Assuming a period of the Mira variable of 527 days \citep{whit87} and using the ephemeris of \citet{yudi94}, they find that the observation were taken at phase 0.65 (i.e. shortly after the photometric maximum) and at phase 1.08, close to the MIRA minimum, respectively.
Both studies consider single and double shell models generated with the DUSTY code. This public domain simulation code solves the problem of radiation transport in a circumstellar dusty environment by integrating the radiative transfer equation in plane-parallel or spherical geometries \citep{ivez96a,ivez99}.
The studies consider different a priori (fixed) input parameters:
\begin{itemize}
\item S01 : L$_{cool}$=5000L$_{\odot}$ and L$_{hot}$=9200L$_{\odot}$ luminosities considering the period-luminosity relation for Miras by \citet{feas96} and the data of \citet{murs97}, respectively. They also take a synthetic stellar spectrum for the Mira from \citet{leje97} with a temperature of 3000K and log$g$=-0.29.
\item B01 : L$_{cool}$=10600L$_{\odot}$ and L$_{hot}$=9000L$_{\odot}$ luminosities considering the period-luminosity relation obtained from observations of Mira oxygen stars in the Large Magellanic Cloud \citep{feas89} and the relation of \citet{murs94}, respectively. They impose that the shell outer boundary associated with the Mira and the WD is situated at thousand times the inner boundary radius. Finally, they take into account a 2600K blackbody spectrum for the Mira.
\end{itemize}

Both authors consider a 2$\times$10$^{5}$ K blackbody for the White Dwarf.
In order to homogeneously compare the SEDs and visibilities of the models, we do not consider the synthetic stellar spectrum used by S01 and used a 3000K blackbody stellar spectrum for the Mira.
The standard size distribution (MRN grain size distribution) as described by \citet{math77}, $n(a)$$\propto$$a^{-3.5}$ with minimum and maximum grain sizes of 0.005 $\mu$m and 0.25 $\mu$m respectively, and a chemical composition of warm silicate dust (W-Sil) with optical properties given by \citet{osse92}, are also assumed by the 2 authors.

\subsection{Single shell models}
\begin{figure*}[tbp]
\centering
\sidecaption
\includegraphics[width=7.5cm]{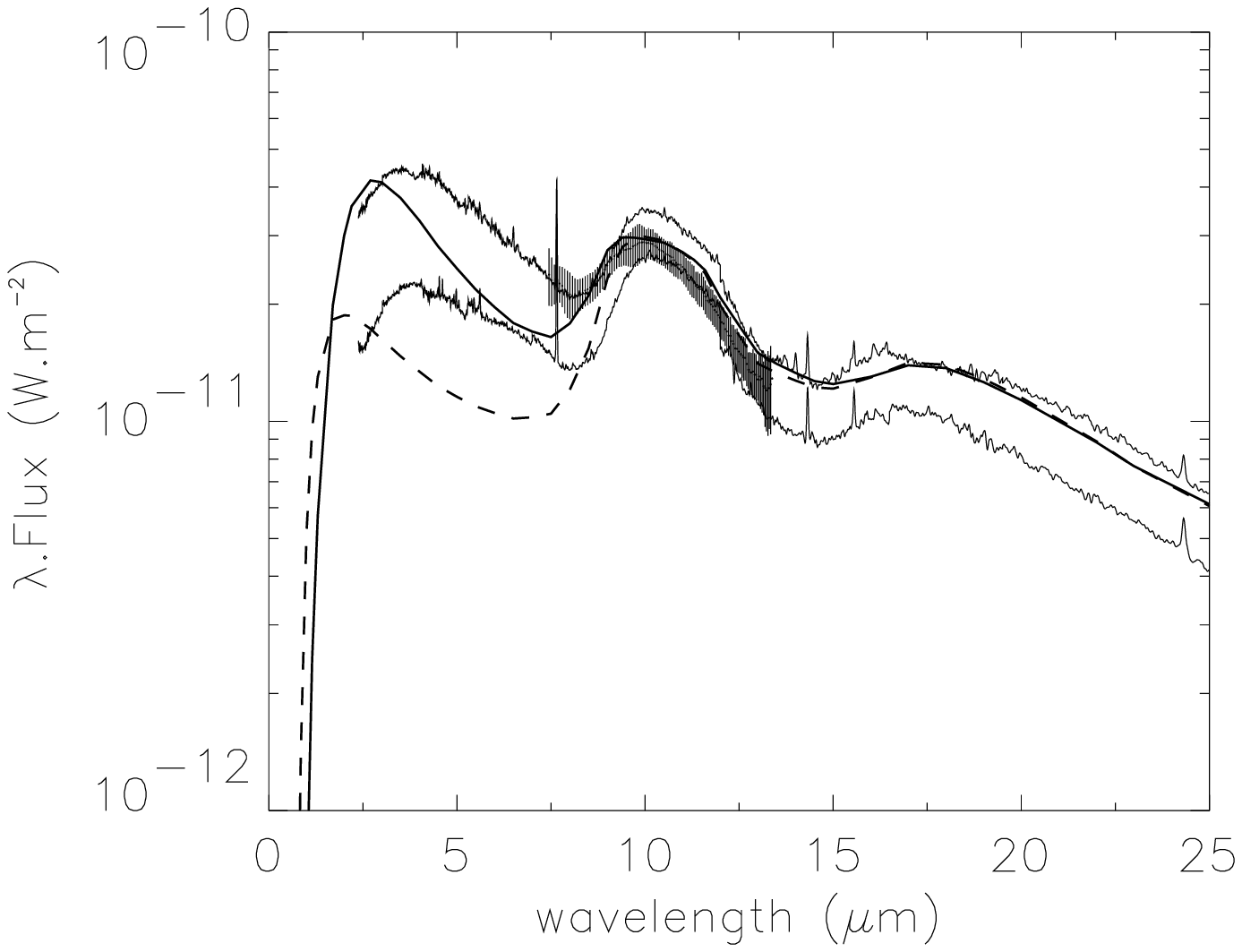}
\includegraphics[width=7.5cm]{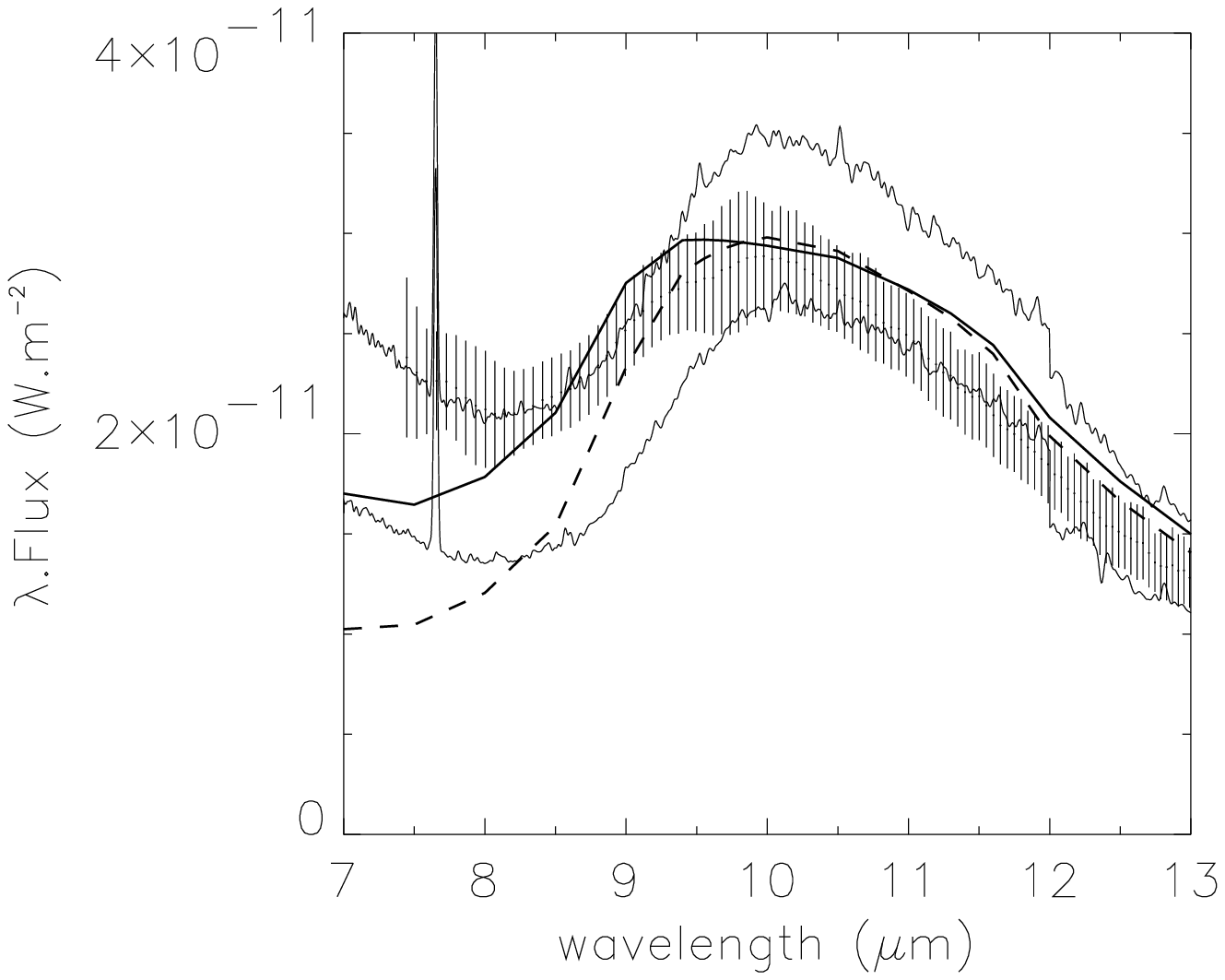}
\caption{Left: best fitting S01 single shell model (thick solid line) and best fitting B01 single shell model (thick dashed line), error bars correspond to the MIDI flux, both thin solid line are related to the ISO/SWS spectra. Right: close-up view of the best-fitted dust feature with the same labels.}
\label{single-shell-SED}
\end{figure*}

The SED of single shell models are shown  in Fig.~\ref{single-shell-SED}. The corresponding dust shell output parameters given by the DUSTY code for S01 and B01 are listed in Tab.~\ref{single shell table}.

Although both models provide good quality SED fits, their parameters are very different. One important divergence comes from the inner boundary temperature of the shell. S01 find a temperature (1600K) close to the condensation temperature of the grains while B01 find a weaker temperature (700K). The inner radius from B01 is 8 times further than the one found by S01. The S01 visual optical depth is about twice larger ($\tau_{V}$=29) than the one found by B01 ($\tau_{V}$=12.5) with a dust density power law coefficient of 2.0 for B01 and 1.8 for S01. As a consequence, the S01 dust shell is more opaque than the B01 one, considering an identical dust chemical composition (warm silicate dust).
B01 are satisfied with their single shell model whereas S01 think that the relatively high optical depth involved with their model ($\geq$30 mag) is not compatible with the JHK photometry by \citet{kama99} suggesting that the light from the Mira is not highly absorbed in near-IR. For this reason, the single shell model is rejected by S01, with the additional argument that such a model is probably simplistic in view of the disturbing presence of the hot companion.

\begin{table*}[tbp]
\centering
\caption{Summary of the single shell DUSTY fit of \object{HM\,Sge} data. Values marked with an asterix were pre-determined and fixed.}
\label{single shell table}
\begin{tabular}
[c]{c|c|c} \hline\hline
 & \citet{schi01} & \citet{bogd01} \\\hline\hline
Effective temperature (K) & 3000$^{*}$ & 2600$^{*}$ \\
Luminosity (L$_{\odot}$)& 5000$^{*}$ & 10600$^{*}$ \\
Distance (kpc) & 1.5 & 2.6 \\
Central star diameter (mas) & 1.6 & 1.8 \\
Grain chemical composition & 100\% W-Sil$^{*}$ & 100\% W-Sil$^{*}$ \\
Grain size distribution & MRN$^{*}$ & MRN$^{*}$ \\
Density power law coefficient & 1.8 & 2.0$^{*}$ \\
Inner boundary temperature (K) & 1600 & 700 \\
Geometrical thickness ($r_{in}$) & 1000 & 1000$^{*}$ \\
Envelope inner radius (mas) & 3.5 & 15.1 \\
Visual optical depth & 29 & 12.5 \\
10$\mu$m optical depth & 2.47 & 1.77 \\ \hline
\end{tabular}
\end{table*}

\subsection{Double shell models}
Two shell models are an attempt to take into account the fact that each stellar component has its own effect on the dust generated in the Mira wind. The 2-component model can physically be justified with the scenario of a strong UV light source placed in the expanding shell of a red giant destroying the dust grains and generating an ionised nebulosity. Radiation pressure from the companion will secondly slow the red giant wind and generate a compression wave in the transition zone in which the conditions for dust formation are very favorable, so that the thin dust rich layer can be identified with its structure. For sake of simplicity, both studies assume that none of the component heats the dust shell of the other and that the total emitted flux is a simple addition of the individual fluxes from each shell. The best parameters found by S01 and B01 are gathered in Tab.~\ref{two shell table}.

\begin{figure*}[tbp]
\centering
\sidecaption
\includegraphics[width=7cm]{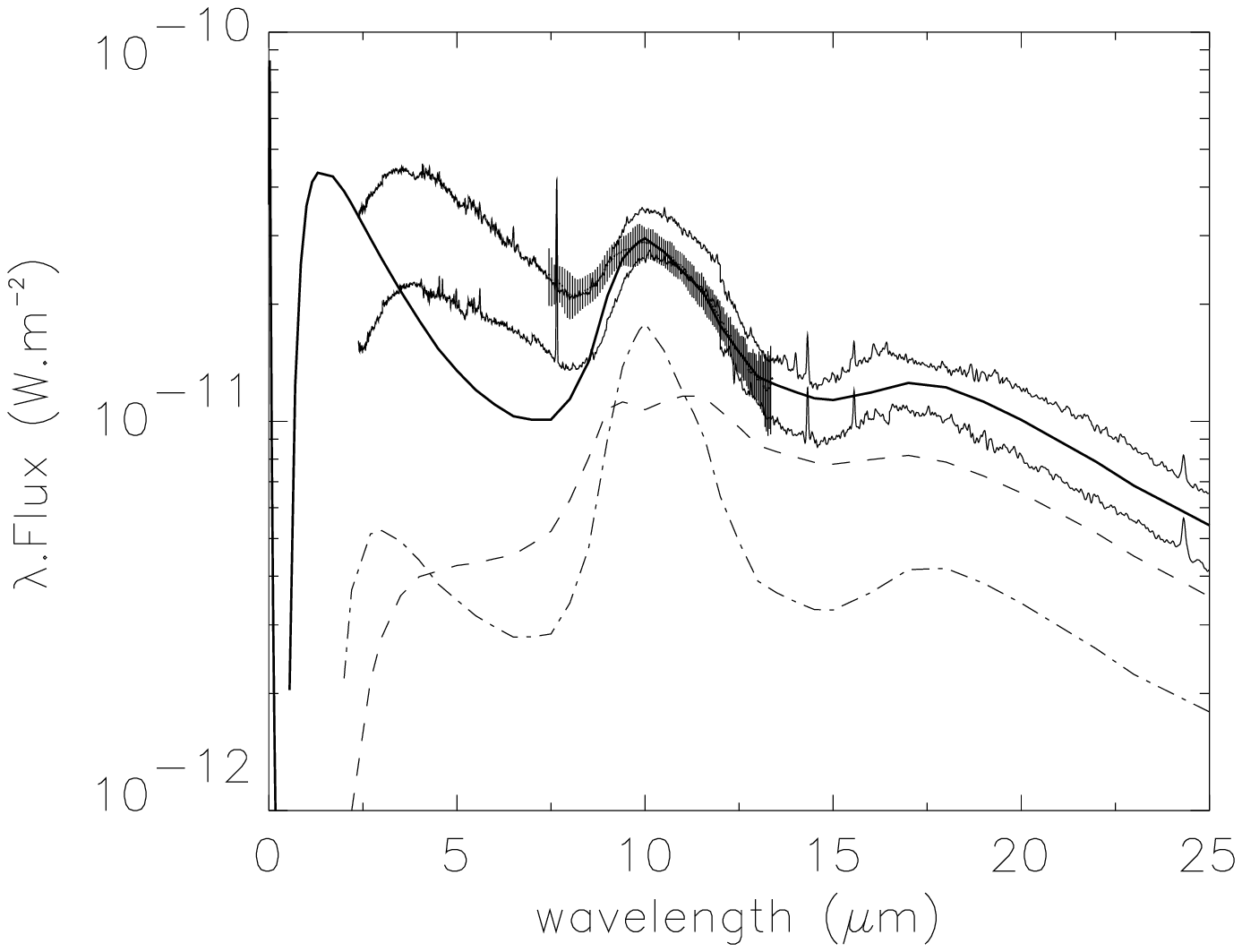}
\includegraphics[width=7cm]{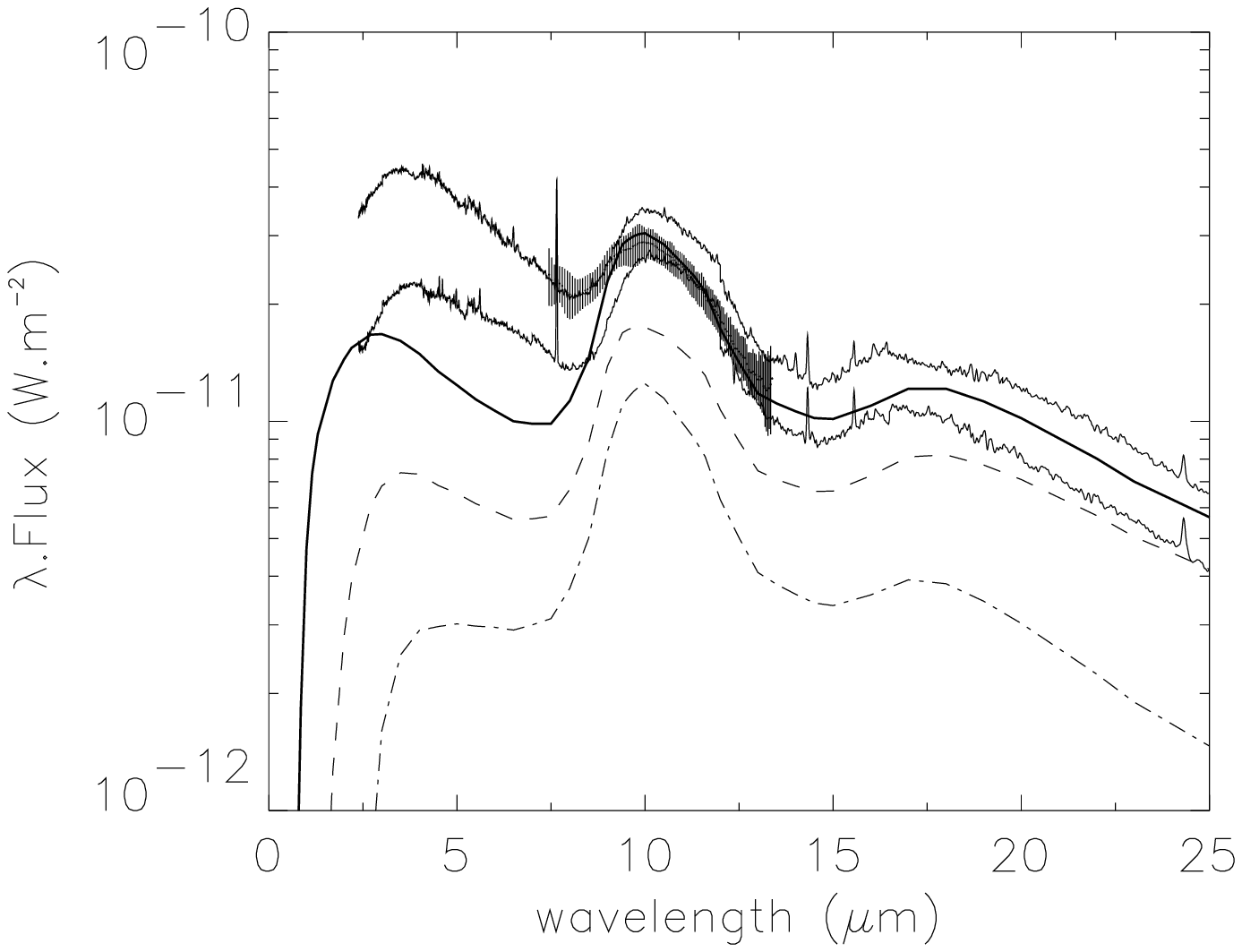}
\includegraphics[width=7cm]{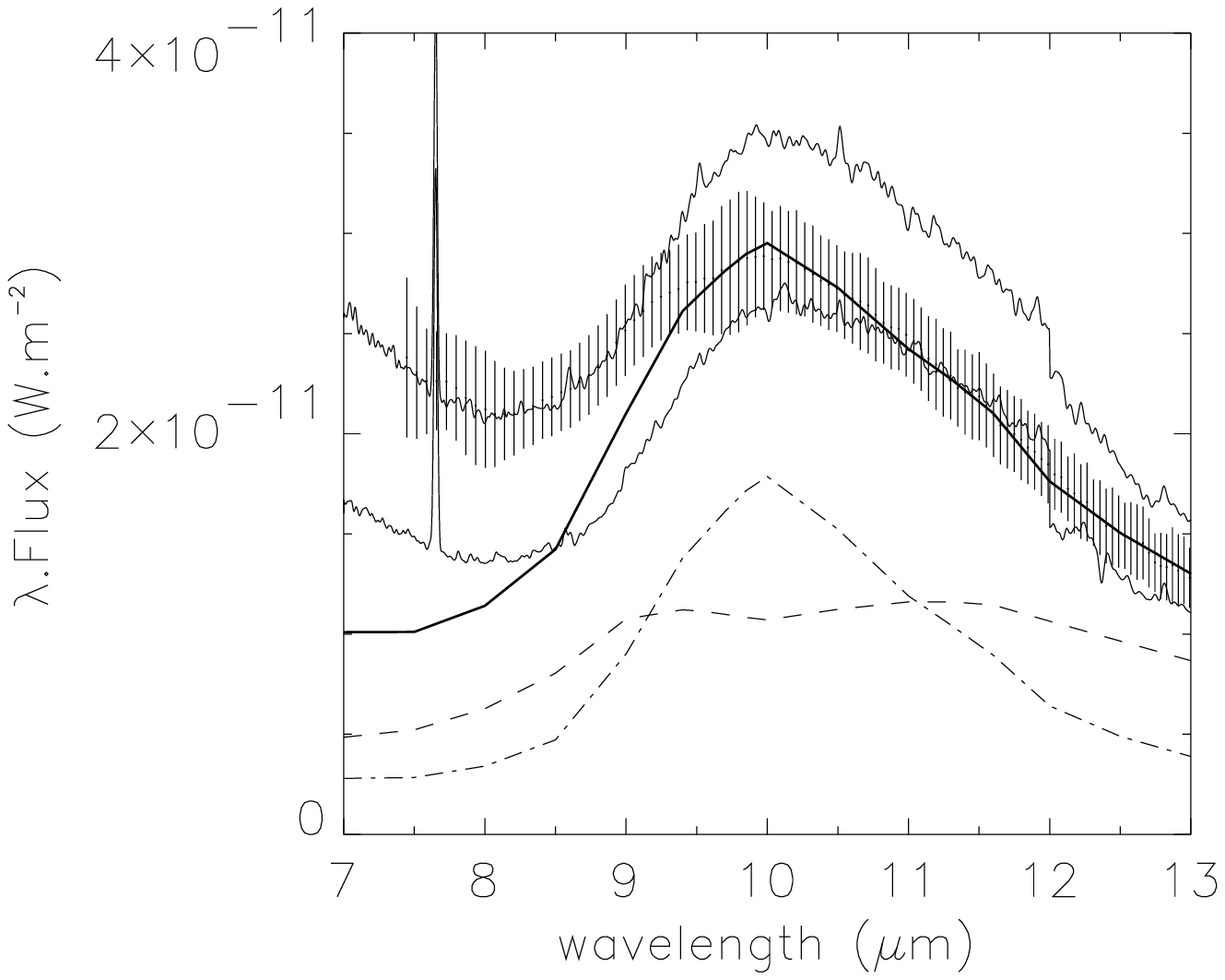}
\includegraphics[width=7cm]{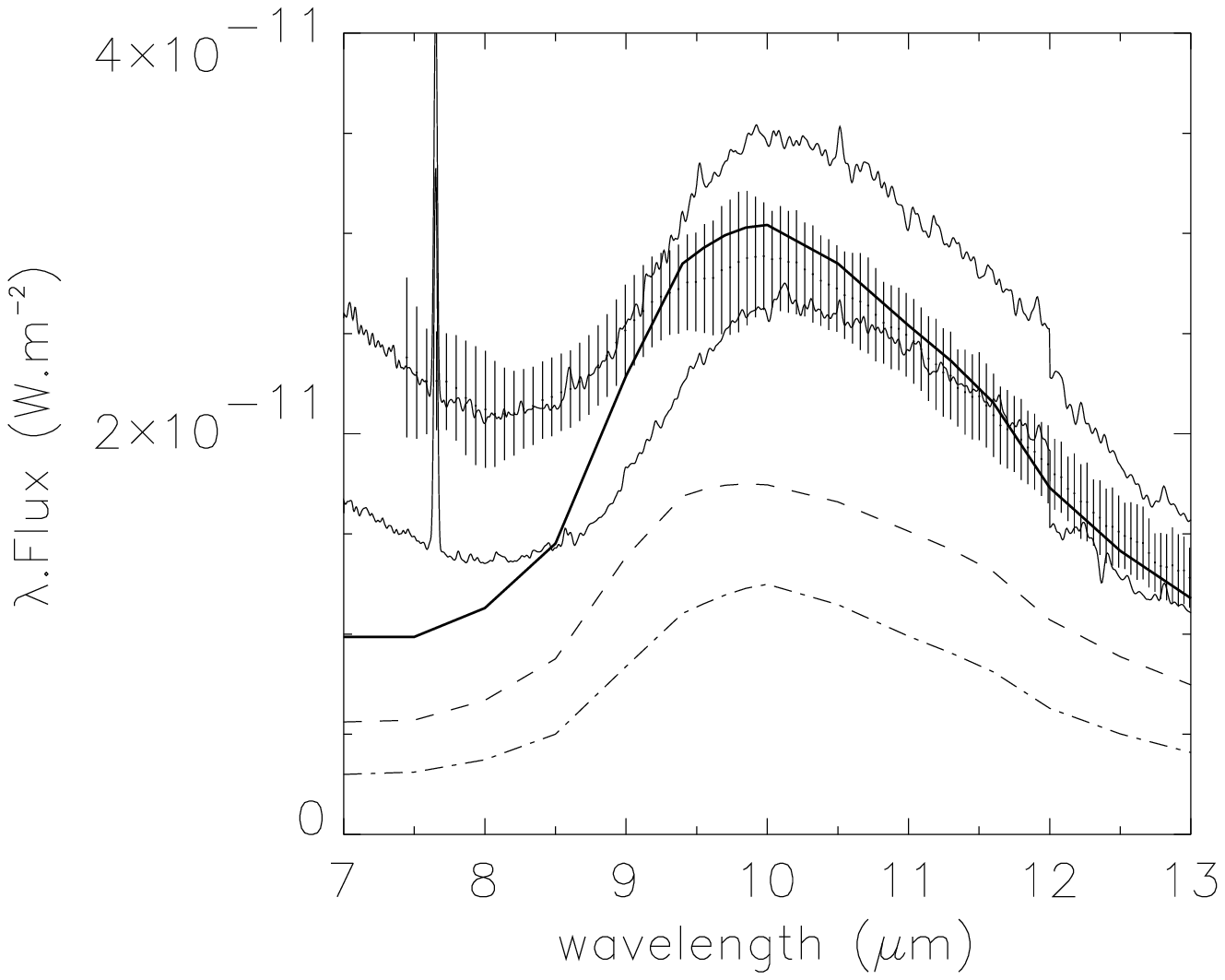}
\caption{Top-left: best fitting S01 two shell model (thick solid line). Top-right: best fitting B01 two shell model (thick solid line), the thin dot-dashed line is the flux contribution of the Shell M (associated with the Mira), the thin dashed line is the flux contribution of the Shell WD (associated with the White Dwarf), error bars correspond to the MIDI flux and both thin solid line are related to the ISO/SWS spectra. Bottom: close-up view of the corresponding best-fitted dust feature with the same labels.}
\label{two-shell-SED}
\end{figure*}

S01 combine the two dust shells with a partition factor $p$ such that the observed flux:

\begin{equation}
\centering
F_{obs}=s(F_{WD Shell}+pF_{Mira Shell})
\end{equation}

where the scaling factor s converts the model flux at the distance of \object{HM\,Sge} into the observed flux.
One of the goals of the two shell modeling by S01 is to take into account the fact that the Mira in \object{HM\,Sge} is optically observable, at least in the near-IR. Hence, they fix an a-priori visual optical depth range between 0.5 and 3.5, much lower than in the single shell hypothesis. The authors also impose that the dust shell thickness associated with the White Dwarf should be geometrically thin (i.e. Y$_{out}$/Y$_{in}$$\leq$5) and therefore could be approximated with a flat density distribution (x=0).
The parameters of this geometrical configuration (Tab.~\ref{two shell table}) consist in an optically thin, geometrically thick shell associated with the Mira, and a geometrically thin, optically thick shell associated with the White Dwarf. The S01 2-component model is found satisfactory by \citet{schi01} but does not fit well shorter wavelengths dominated by the Mira spectrum. This can be attributed to the variability of the Mira which is not taken into account by the model.
Despite some discrepancies, the authors favor the 2-component models approach.
Fig.~\ref{two-shell-SED} shows their two shell best fitted model. The top-left part shows the S01 spectral energy distribution (SED) with the flux contribution of the individual shell, while its bottom-left part shows a close-up of the silicate dust peak around 10$\mu$m.

The two shells model of B01 does not use any partition factor: it assumes that the flux from the central source is equal to the sum of two blackbodies located within a sphere of radius $r_{in}$. They argue that, given the small distance between the components (i.e. 40\,mas) compared to the size of their common dust envelope, this condition is approximatively satisfied. 
Contrary to the conclusion of S01, the results of the B01's study given in Tab.~\ref{two shell table} lead the authors to conclude that the 2 shell model is unable to describe the observed energy distributions at mid-IR wavelengths due to a twice larger squared deviation comparing to the single shell model. Moreover, the distance is a free parameter of the model and B01 are lead to conclude that the overestimated distance values of the best fits (3.7kpc) is in contradiction with previous studies of \object{HM\,Sge}. 
Fig.~\ref{two-shell-SED} shows their two shell best fitted model. The top-right part of Fig.~\ref{two-shell-SED} shows the B01 spectral energy distribution (SED) with the flux contribution of the individual shell, while its bottom-right part shows a close-up of the dust peak around 10$\mu$m.

\begin{table*}[tbp]
\centering
\caption{Summary of the two shell DUSTY fit of \object{HM\,Sge} data. Shell "M" is the dust shell associated with the Mira component, Shell "WD" is the dust shell associated with the White Dwarf component. Values marked with a star were pre-determined and fixed.}
\label{two shell table}
\begin{tabular}
[c]{c|cc|cc} \hline\hline
 & \multicolumn{2}{c|}{\citet{schi01}} & \multicolumn{2}{c}{\citet{bogd01}} \\\hline\hline
 & M & WD & M & WD \\\hline
Effective temperature (K) & 3000$^{*}$ & 2$\times$10$^{5}$$^{*}$ & 2600$^{*}$ & 2$\times$10$^{5}$$^{*}$ \\
Luminosity (L$_{\odot}$)& 5000$^{*}$ & 9200$^{*}$ & 10600$^{*}$ & 9000$^{*}$ \\
Distance (kpc) & 1.5 & 1.5 & 3.7 & 3.7 \\
Central star diameter (mas) & 1.6 & 5$\times$10$^{-4}$ & 1.3 & 2$\times$10$^{-4}$ \\
Grain chemical composition & 100\% W-Sil$^{*}$ & 100\% W-Sil$^{*}$ & 100\% W-Sil$^{*}$ & 100\% W-Sil$^{*}$ \\
Grain size distribution & MRN$^{*}$ & MRN$^{*}$ & MRN$^{*}$ & MRN$^{*}$ \\
Density power law coefficient & 1.6 & 0$^{*}$ & 2$^{*}$ & 2$^{*}$ \\
Inner boundary temperature (K) & 1400$^{*}$ & 800 & 900 & 900 \\
Geometrical thickness ($r_{in}$) & 1000 & 1.1 & 1000$^{*}$ & 1000$^{*}$ \\
Envelope inner radius (mas) & 3.3 & 128.7 & 6.1 & 40.0 \\
Visual optical depth & 1.5 & 13 & 8.8 & 8.8 \\
10$\mu$m optical depth & 0.13 & 1.11 & 0.75 & 0.75 \\
Relative contribution & 10 & 1 & 1 & 1 \\ \hline
\end{tabular}
\end{table*}

\bigskip

We reproduce the models of \citet{schi01} and \citet{bogd01}. It is very interesting to point out that the values of the parameters and their respective conclusions are in complete opposition, B01 favor a single shell model while S01 favor a two shell model. The extensions of the dusty shells are very different from one model to another, and our mid-infrared interferometry data bring an important constraint that can help us to disentangle between them.

\section{Visibilities of the models}
\label{sect:Vismodels}

\subsection{Description}

In this section, we use DUSTY to generate the synthetic spectrally-dispersed (7.5-13$\mu$m) visibility profiles for the 2 single shell and 2 double shell models of S01 and B01 for each projected baseline of the MIDI observations. Then we compare the synthetic visibilities with the mid-infrared visibility measurements recorded. More details on the computation of the synthetic visibilities with the DUSTY code can be found in Appendix A.

\subsection{Single shell visibility models}

\begin{figure}
\includegraphics[width=9cm]{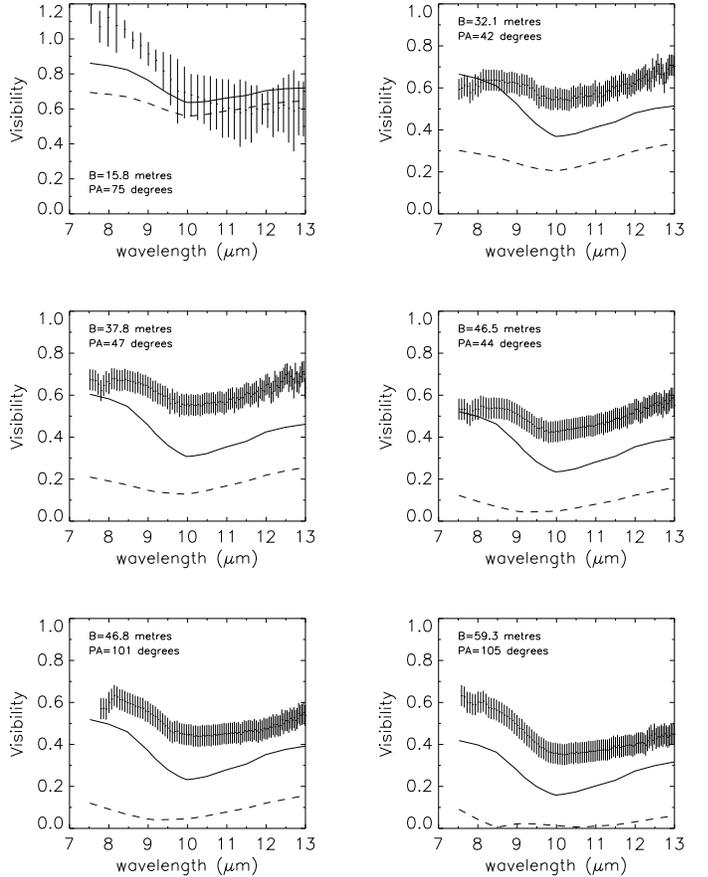}
\caption{Corresponding S01 (solid line) and B01 (dashed line) single shell visibility models superimposed on the MIDI visibilities (error bars) for the six projected baselines.}
\label{single-shell-visibility}
\end{figure}

Figure \ref{single-shell-visibility} shows the S01 (solid line) and B01 (dashed line) single shell visibility models superimposed on the MIDI visibilities (error bars) for each projected baseline. These plots show that, whatever the model, the visibilities are under-evaluated by comparison with the MIDI data. Nevertheless, the S01 model is relatively satisfactory, close to the observed visibilities, with a similar spectral shape. The B01 model provides much lower visibilities.
We point out that the visibility curve shape of the S01 model is purely related to the chemistry of the shell and not a consequence of the geometry of the source. The drop of visibility at 10$\mu$m is the consequence of the silicate emission, leading to a more extended object at these wavelengths \citep{ches05a}. By contrast, the shape of the visibility curves from the B01 model depends on the projected baseline with the wavelength of minimum visibility changing from 9.2 to 10$\mu$m. The shape of the B01 visibility curve reflects the change of angular diameter due to the silicates, but also the modulation of the visibility due to the ring resolved by the interferometer. The inner radius of the B01 model is large (angular diameter of 30\,mas), and the dust shell is rather optically thin ($\tau_{10\mu m}$=1.8). Therefore, the dust emission is seen dominantly as a ring at low visibility level, whereas the S01 model is mostly seen as a compact source. Such modulations from an extended dusty ring have already been observed with MIDI \citep{ches06}.

The shape of the observed visibility curves is relatively constant, because the shell is optically thick or because the inner rim radius is too small to be resolved by the interferometer. The S01 model fulfills these two conditions: the 10$\mu$m optical depth of the shell is 2.5, and the angular diameter of the inner rim is only 7\,mas.

Overall, the S01 model is impressively good, keeping in mind that the SED fits are usually not unique, and do not provide good predictions for the observed radii of dust emission. 

\subsection{Double shell visibility models}
Computing the contribution from two separate shells and adding them to form a resulting SED with the DUSTY code is a simple operation provided that the fluxes are supposed to be independent. This strong hypothesis is not viable in the context of our interferometric observations that are also sensitive to the relative position of the two shells. Hence, it is necessary to add to the description of the two shells their projected separation and position angle of their centers. Several options are possible: we position the centers of the two shells at the same position, or we position the centers at a given position and consider them as fixed parameters, or we consider these news parameters as free and part of the fitting process. In order to limit the extent of the current study, we fixed the parameters to the values of the binary found in \citet{eyre01}: the Mira and WD shells are positioned 40\,mas apart at a PA of 130$^\circ$. The detailed visibility computation is given in Appendix B.

\begin{figure}
\includegraphics[width=9cm]{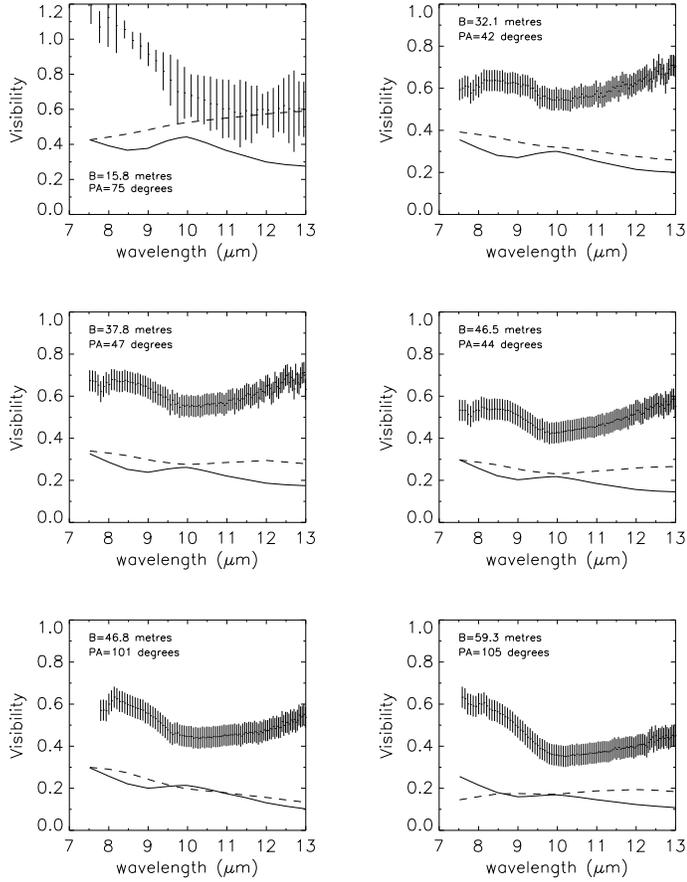}
\caption{Corresponding S01 (solid line) and B01 (dashed line) double shell visibility models superimposed on the MIDI visibilities (error bars) for the six projected baselines.}
\label{double-shell-visibility}
\end{figure}

The direct consequence of the introduction of an extended shell in the two components models is a significant decrease of the N band visibilities of the S01 and B01 model. In addition, some differences in the shape of the visibility curves can be noticed (see Fig.\ref{double-shell-visibility}). The shapes of the visibilities of the B01 model exhibit a strong dependency on the position angle of the baseline. When the orientation of the projected baseline ($\sim$45$^\circ$) is perpendicular to the binary separation, i.e. the projected angular separation between the Mira and the WD is small ($\sim$3.5\,mas for PA$\sim$46$^\circ$) compared to the spatial resolution of the interferometer (Fig.\ref{bina_sepa}), the spectral shape and the level of the visibilities are naturally explained by the predominant correlated flux from the Mira shell seen by the interferometer. The unique effect from the extended (over-resolved at baselines larger than 30m) and optically thin ($\tau_{10\mu m}$=0.75) WD shell is to dilute the correlated flux to a low level with little spectral signature. When the orientation of the projected baseline gives a larger projected angular separation between the Mira and the WD ($\sim$23-35\,mas at PA$\sim$75-100$^\circ$) compared to the spatial resolution of the interferometer (Fig.\ref{bina_sepa}), the shift between the Mira and WD shells in the sky introduces an interferometric signal that mimics the one from a binary. Because the interferometer is able to resolve the 2 shells with their own photocenter, a variation occurs on the slope of the visibility depending on the projected baseline. The binary effect is given by a sinusoidal dependency of the visibility shape versus the projected angular separation of the 2 components. The visibility increases from 8 to 13$\mu$m for the 15.8m baseline, then decreases for the 46.8m one, and increases again for the 59.3m one. This binary signal is not visible in the S01 model since the WD shell is over-resolved whatever the projected baseline and do not contribute to the correlated flux. The other differences between the shape of the S01 and B01 visibility curves come from the flux balance between the emission of the two shells at various wavelengths, as can be seen in Fig.\ref{two-shell-SED}. The ratio of the two fluxes do not vary in the B01 model whereas a steep variation occurs between 9 and 11$\mu$m in the S01 model. At 9.7$\mu$m, the Mira shell dominates the flux leading to a local increase of the visibility.

Clearly it is debatable whether these dust shell models associating two {\it independent} flux sources and dust distributions are good approximations of complex sources as symbiotic novae and provide a better way of understanding them. More complex geometries should be used, constrained by hydrodynamical models and high resolution imaging, as a further step for improving our knowledge of these sources.

\section{S01 single shell model improvement}
\label{sect:modelImp}

In this section we consider the S01 single shell model as the potentially best model able to account for spectra and visibilities. We propose a pragmatic strategy of improvement that consists in small perturbations of the already satisfactory parameters of the S01 single shell model. We can see that the model is slightly too extended so that we need to spatially concentrate the emission of the dust shell.

A solution consists in playing with some intrinsic parameters of the dusty environment of \object{HM\,Sge}, i.e. the density distribution power law and the optical depth. Steepening the density power law (at constant mass) increases the dust density at the inner boundary with the density dropping faster. The emission comes from a shell located closer to the inner boundary and, consequently, less resolved by the interferometer. A decrease of the optical depth (by decreasing the total amount of mass) leads to a more transparent dusty environment, so that deep seated layers contribute significantly to the correlated flux seen by the interferometer. These two changes do not affect the SED in the same way. The dust density distribution can only be constrained from the SED beyond 10$\mu$m since the matter distribution mainly determines the slope of the redistributed energy by dust emission \citep{lobe99}. On the other hand, a variation of the optical depth affects the whole shape of the curve, in particular the silicate feature. However, playing with both parameters, we find that the dust density distribution law has much more influence on the visibilities than a significant decrease of the optical depth. Moreover a decrease of the optical depth changes the whole SED curve that quickly does not fit the ISO and MIDI fluxes anymore. Diminishing the density power law coefficient increases the visibility but differences between the flux data and the model arise at longer wavelengths, with a too small flux of the model compared to the ISO data. It could be argued that this is due to the appearance of a beam size effect toward wavelengths larger than 15$\mu$m, the ISO telescope gathering light from a large field-of-view. But the MIDI visibilities suggest a compact dusty environment that easily fits into the 8m telescope beam of 300\,mas and the ISO emission is most probably also spatially compact at larger wavelengths. 

The perturbed model has a density power law coefficient of 2.2 instead of 1.8 and a 10$\mu$m optical depth of 2.15
instead of 2.47 with the remaining parameters equal to those of the S01 single shell (see Tab.~\ref{single shell table}). The fits are rather good but the optical depth of the perturbed model remains high. This model is subject to the same critics than for the S01 one: the shell prevents the flux from the Mira to be really detectable in near-IR. We tested a last change in order to improve the quality of the fits between 8 and 10$\mu$m. One can notice that the model visibility slope between 8 and 10$\mu$m is too strong revealing a too much extended structure at these wavelengths. Introducing a contribution from an optically thinner material than silicate such as carbon may allow to decrease the size of the object at these wavelengths. 

\begin{figure*}[tbp]
\centering
\sidecaption
\includegraphics[width=7cm]{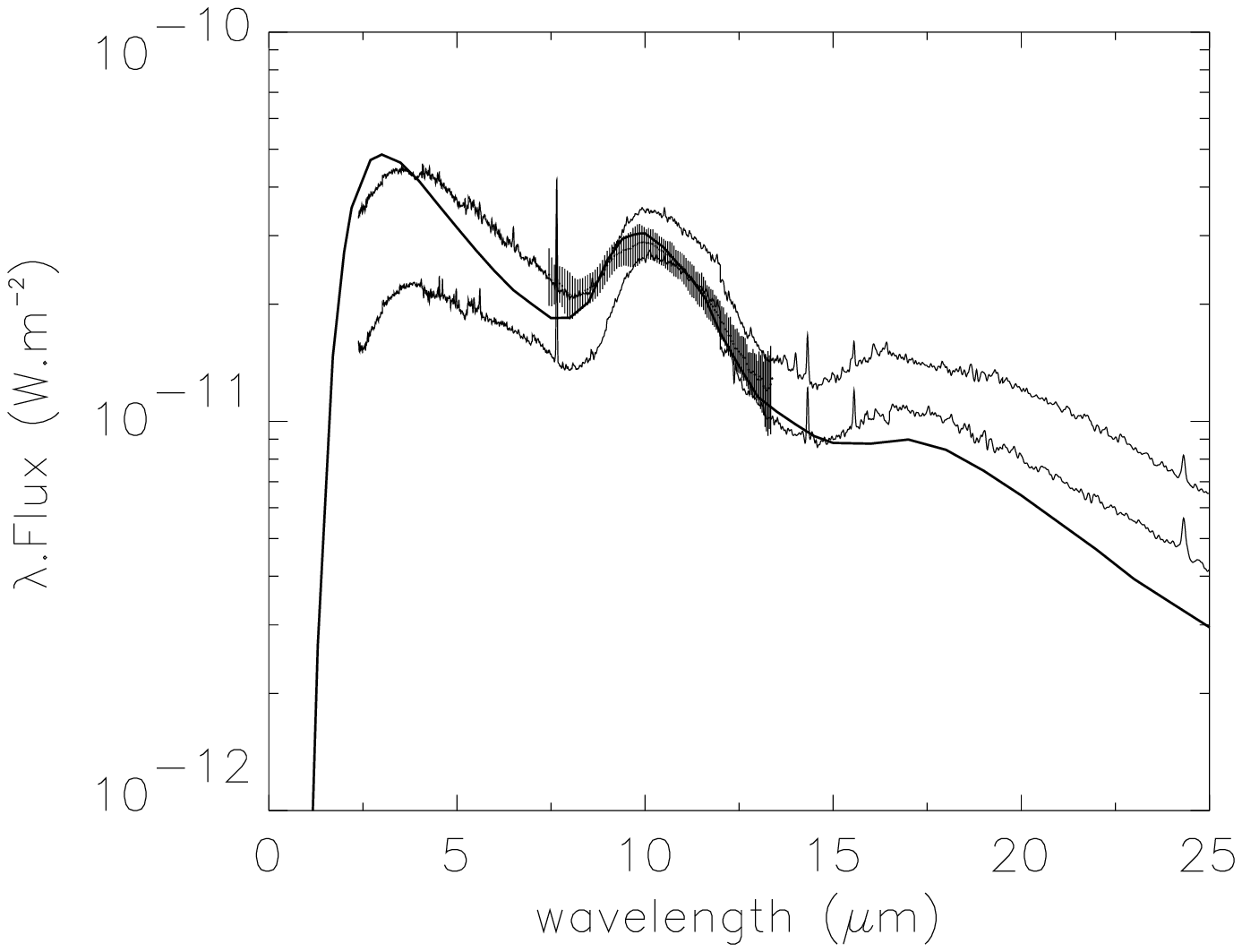}
\includegraphics[width=7cm]{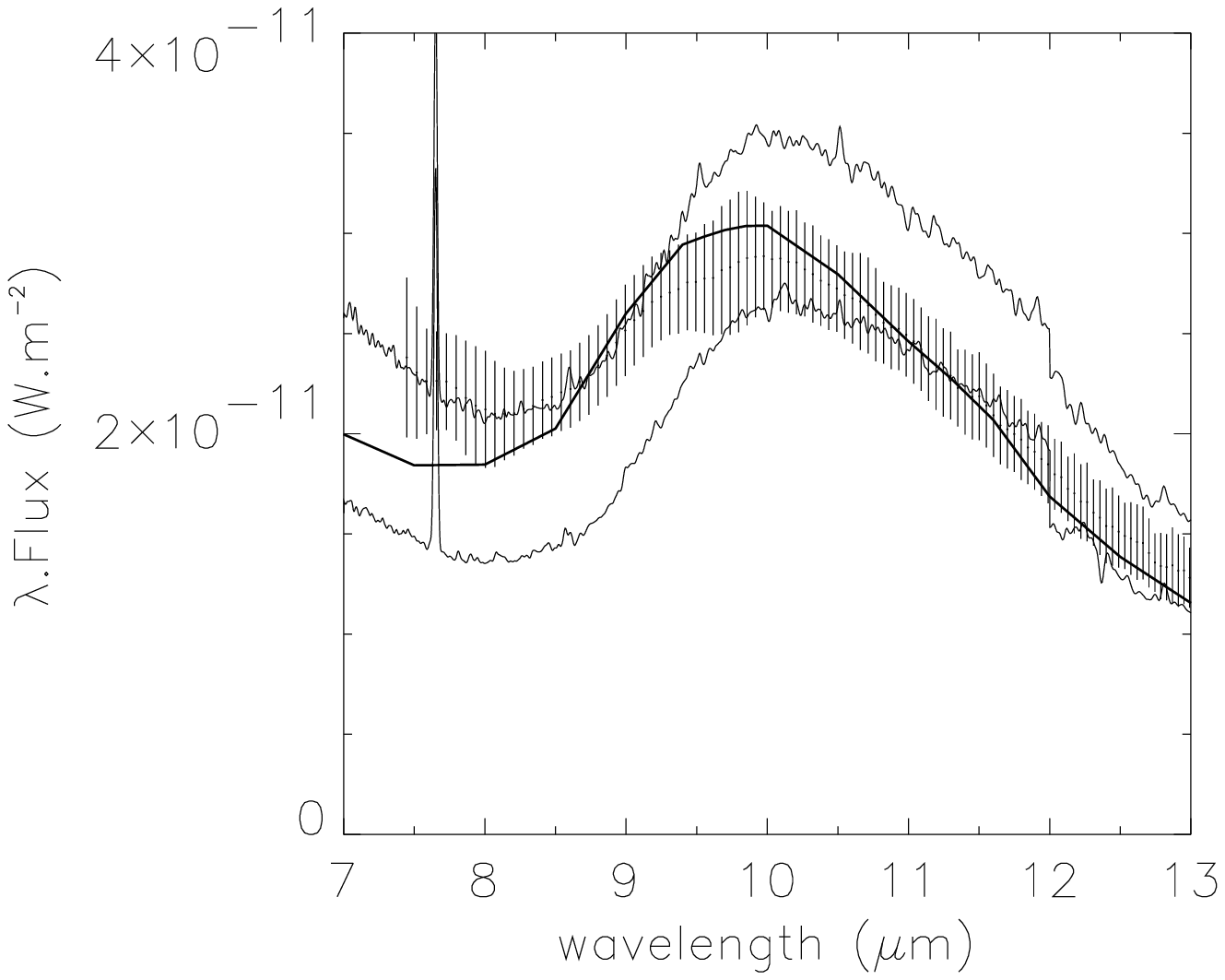}
\includegraphics[width=11cm]{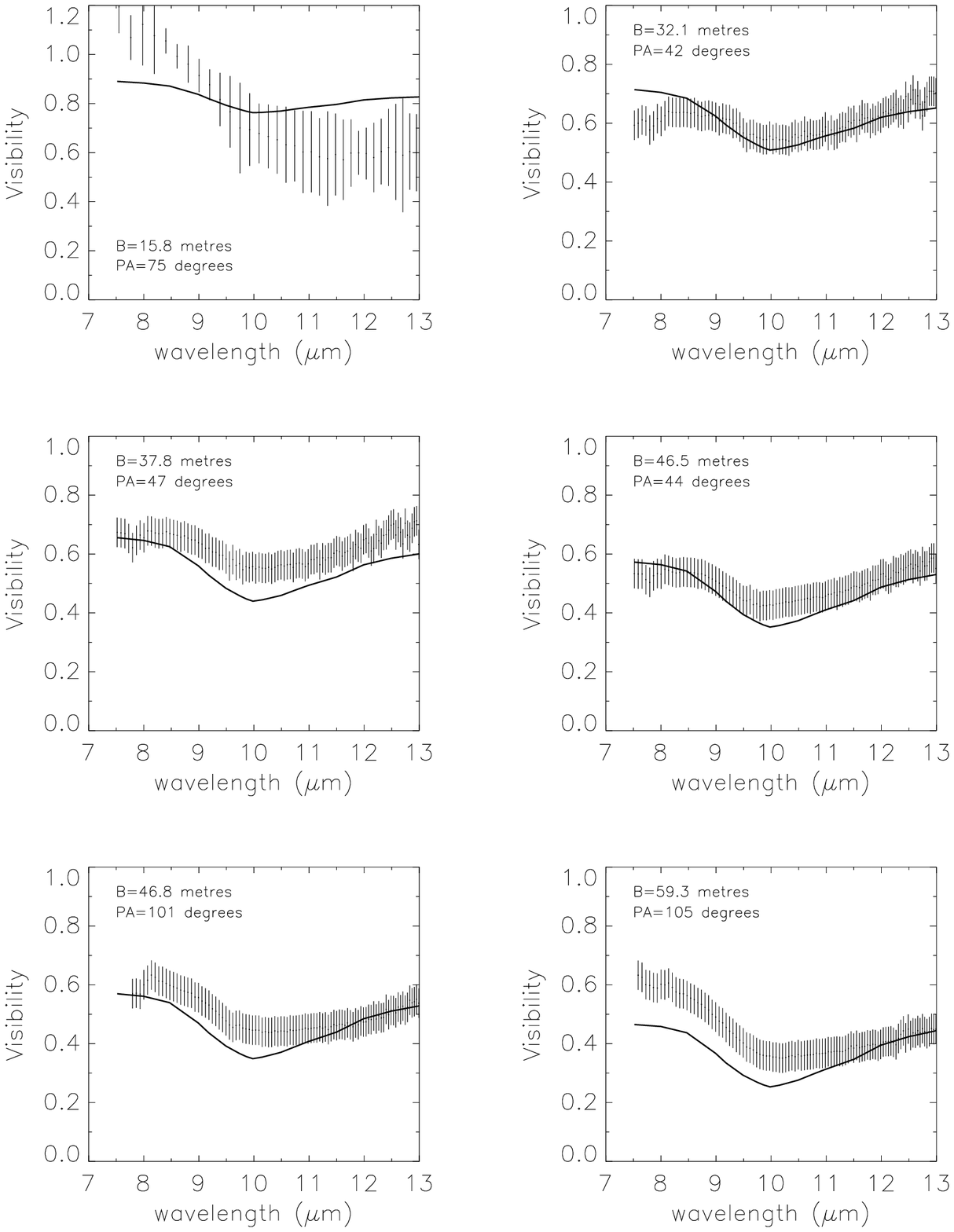}
\caption{Top-left: best fitting of the perturbed S01 single shell model (thick solid line), error bars correspond to the MIDI flux and both thin solid line are related to the ISO/SWS spectra. Top-right: close-up view of the best-fitted dust feature with the same labels. Bottom: Corresponding single shell visibility model superimposed on the MIDI visibilities (error bars) for the six projected baselines.}
\label{perturbation}
\end{figure*}

\begin{figure}
\includegraphics[width=9cm,height=5cm]{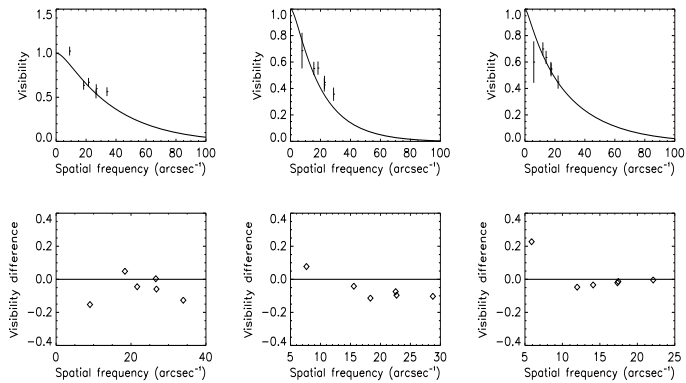}
\caption{Top: Best perturbed visibility profile model superimposed on the MIDI visibilities at 8.5, 10 and 13$\mu$m (from the left to the right). Bottom: Corresponding visibility difference between the model and the MIDI data.}
\label{visibility-comparison}
\end{figure}

\begin{figure}
\centering
\includegraphics[width=9cm]{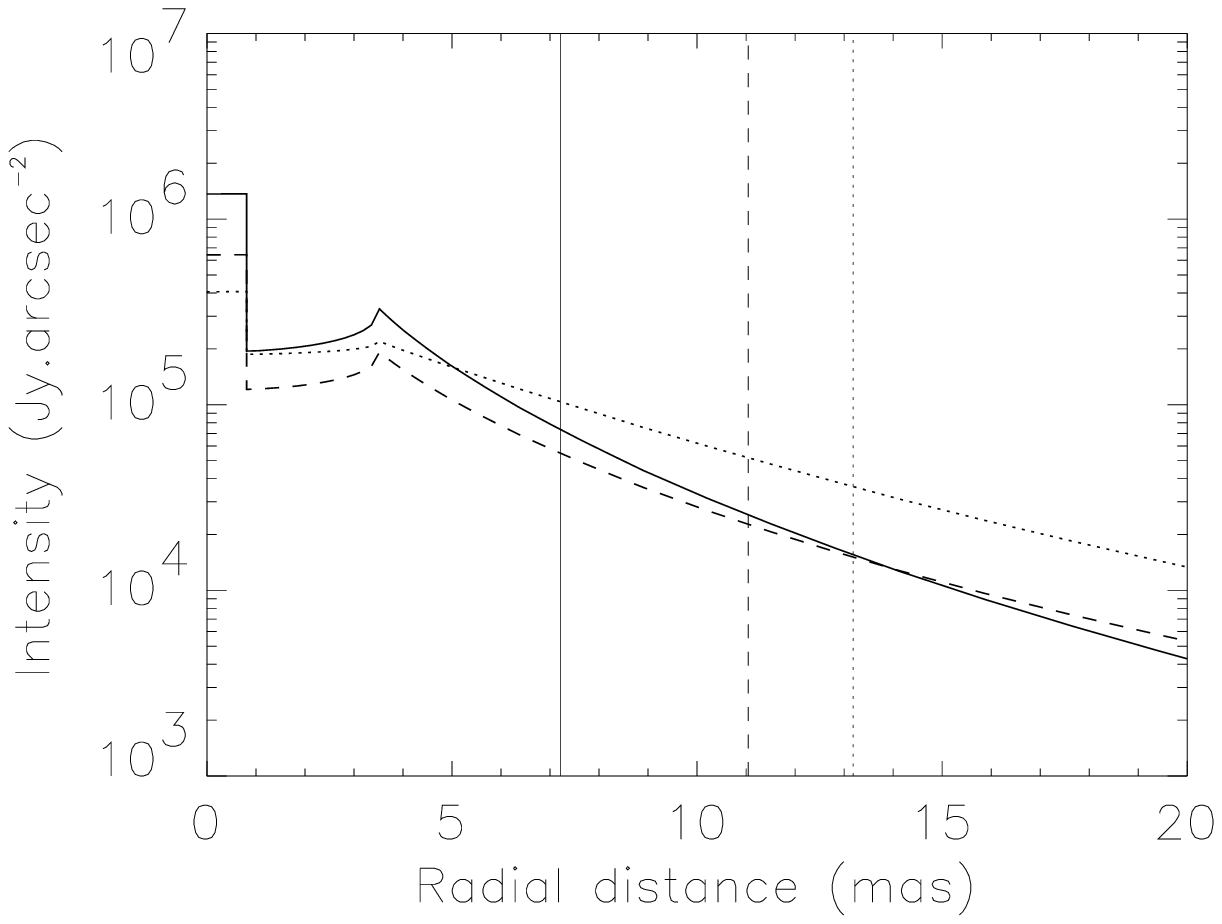}
\includegraphics[width=2.8cm]{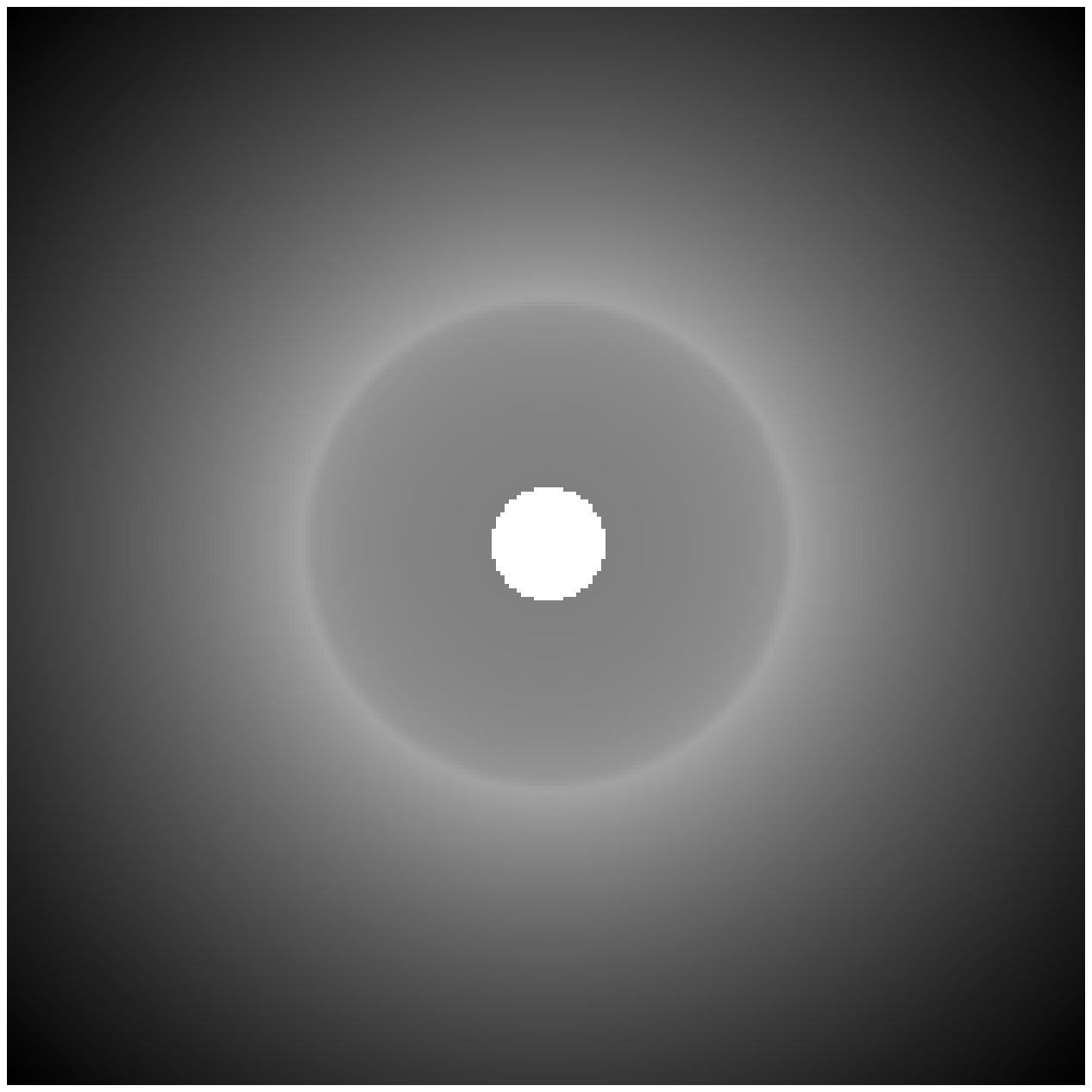}
\includegraphics[width=2.8cm]{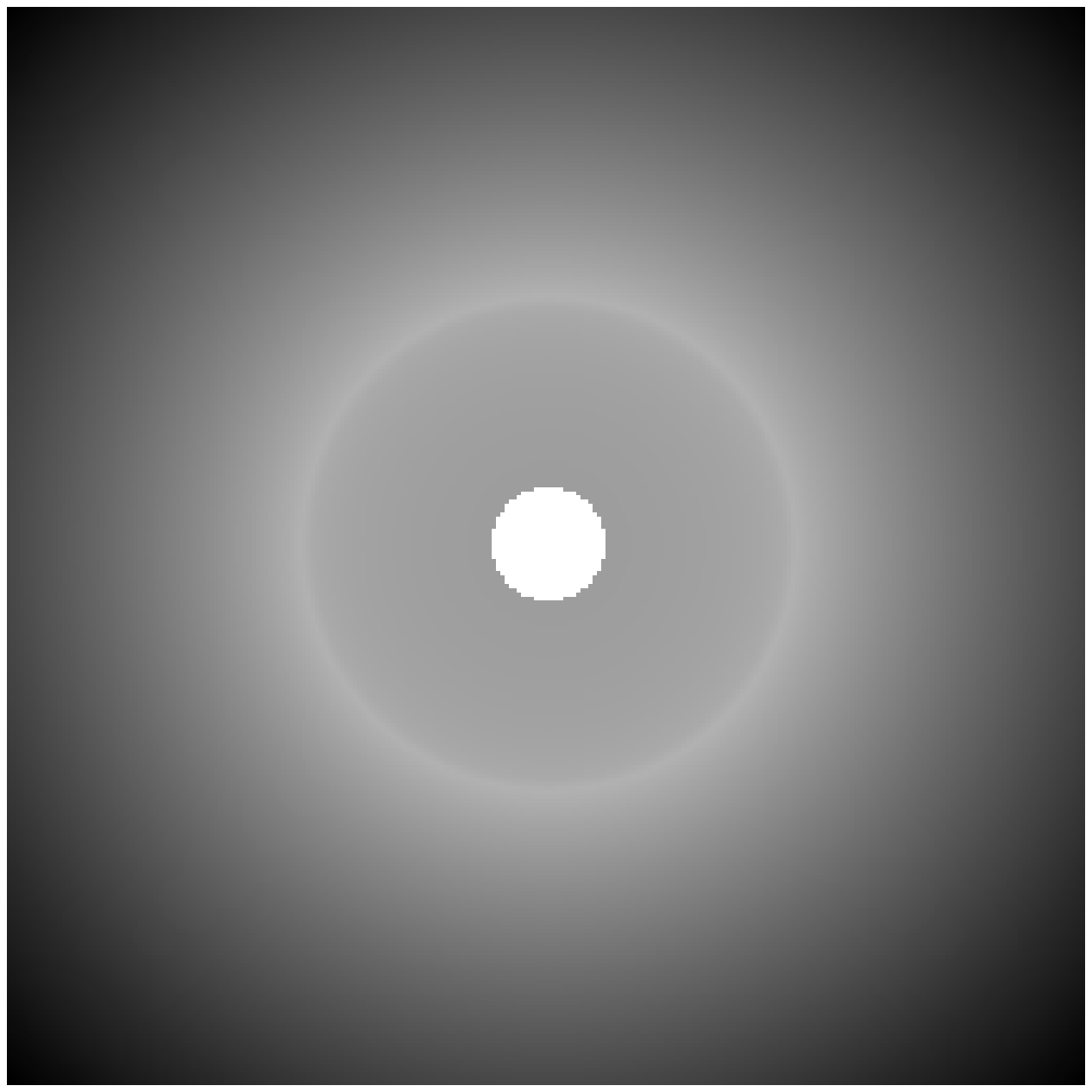}
\includegraphics[width=2.8cm]{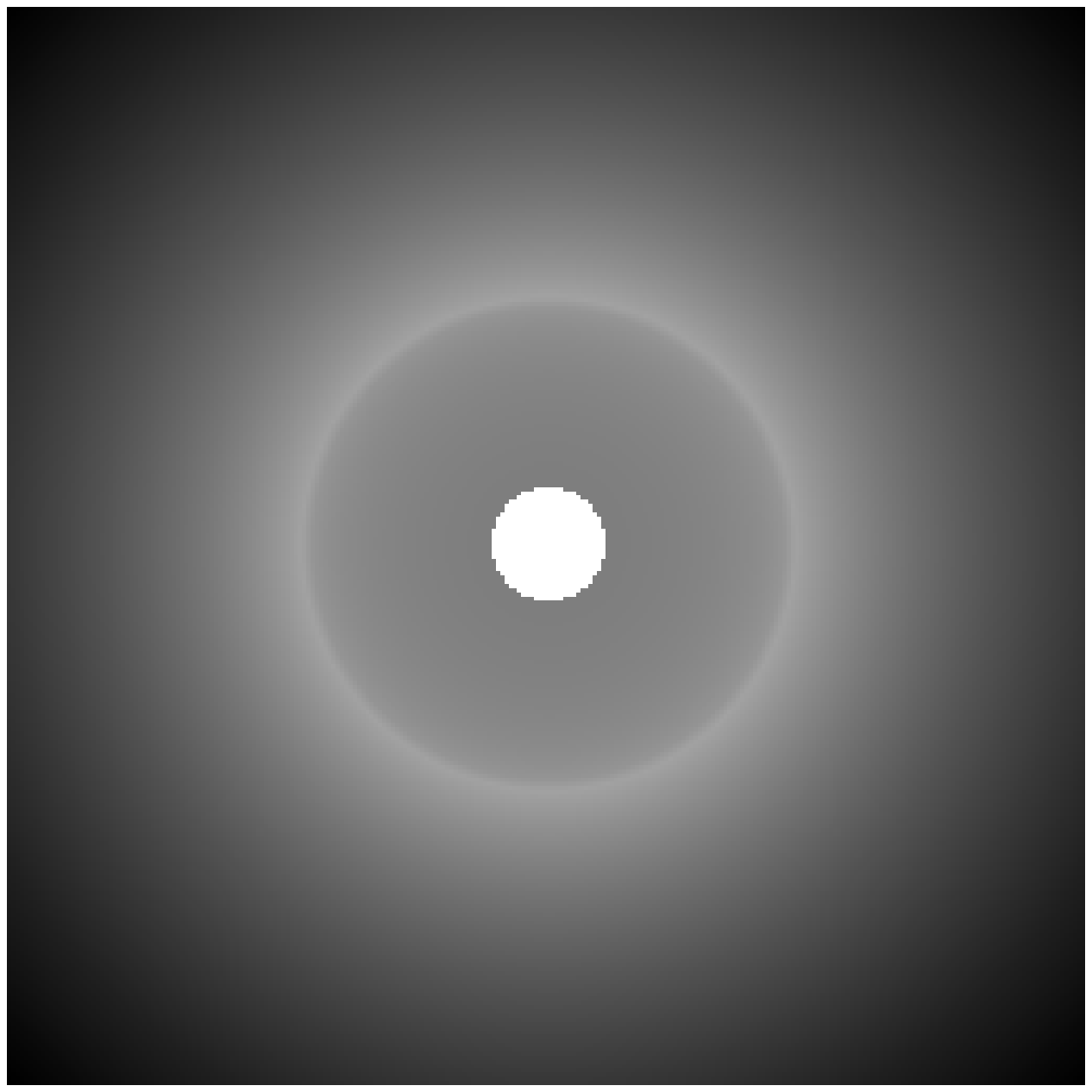}
\caption{Top: Best perturbed intensity profile model at 8.5 (solid line), 10 (dotted line), and 13$\mu$m (dashed line). The vertical lines indicate the HWHM (at 8.5 (solid line), 10 (dotted line) and 13$\mu$m (dashed line)) of the best perturbed intensity profile using a Gaussian fitting. Bottom: The mid-infrared images (at 8.5 (left), 10 (middle), and 13$\mu$m (right)) predicted by the best perturbed model given in a logarithmic scale.}
\label{Intensity}
\end{figure}

Carbon in the form of graphite has been often involved to construct the early SED models \citep{kwok79}, and its presence was justified by the fact that this material would come from the White Dwarf ejecta during the outburst of 1975 or due to particular dust nucleation processes that occur under the non-equilibrium physical conditions that prevail around the WD. Another argument is also provided by \citet{brya91} in their analysis of the time variable 0.22$\mu$m feature in IUE spectra that they attribute to carbon-based material (see also \citet{muna89}).
If present, this carbon contamination must be small and we developed a test model in which the dust composition is a mixture of 90$\%$ of warm silicates \citep{osse92} and 10$\%$ of amorphous carbon \citep{hann88}.
Fig.~\ref{perturbation}, shows the best perturbed S01 single shell model with the new mixture: the density power law coefficient is 1.8 and the 10$\mu$m optical depth has decreased to 1.89.  The other parameters of this model keep the same values of the S01 model contained in Tab.~\ref{single shell table}. Fig.~\ref{visibility-comparison} shows the best perturbed visibility profile model superimposed on the MIDI visibilities at 8.5, 10 and 13$\mu$m and the visibility difference between the model and the MIDI data. Fig.~\ref{Intensity} shows the corresponding best perturbed intensity profile model at the same wavelengths.

\section{Discussion}
\label{sect:discuss}

\subsection{Spherical models: single versus double shells}
The main result of the MIDI observations is that the dust is spatially concentrated around the Mira star and that the MIDI single-dish fluxes is directly comparable to the ISO one. 
Using a Gaussian model, we estimated the HWHM of the dust in the 8-9$\mu$m and 11-12$\mu$m range to 7.8$\pm$1.3\,mas (12AU) and 11.9$\pm$1.3\,mas (18AU) respectively.

We compared the MIDI data with the predictions of various single and double spherical shells spectrophotometric models and favored the single shell models as they provided the most spatially concentrated emission. 
Gaussian model fitting of our best model provide the following Half Width at Half Maximum: 7.2\,mas at 8.5$\mu$m, 13.2\,mas at 10$\mu$m, and 11\,mas at 13$\mu$m (see Fig.~\ref{Intensity}). These extensions are much smaller than the separation between the Mira and the WD of 40\,mas (60AU) determined by the HST observations \citep{eyre01}. At 20\,mas, the local intensity of the envelope is less than 3\% than the one at 4\,mas, and the amount of dust that is located in the vicinity of the WD is limited.

The mass-loss rates of the best S01 and the perturbed models are 7.7$\times$10$^{-6}$ and 7$\times$10$^{-6}$M$_\odot$/yr, respectively. Similar estimations can be found in \citet{rich99}, \citet{nuss90}, and \citet{keny89}.

The difficulty encountered in the fitting process is that a large amount of dust is required to explain the 10-100$\mu$m flux and that the Mira photosphere should be visible in the near-IR. The solution proposed by \citep{brya91} is to use a large inner radius, 10.5R$_*$ in the case of \object{HM\,Sge}, but such a large inner radius would increase strongly the mid-IR extension of the system and is not compatible with the data. In the other hand, the best perturbed S01 model is able to fit many observations in first order but is still far from providing a satisfactory fit of the near-IR and long wavelength SED and is also too optically thick. 

At this point, the main assumption of spherical symmetry has probably reached its limits. Although there is considerable evidence for departures from spherical symmetry at large scales ($\geq$1$\arcsec$), the primary constraints on the models from the MIDI mid-IR visibilities and spectra are not sufficient to warrant more complex models at small scales. The use of more complex models, like the one developed in \citet{lope97} would help us to better constrain the dust density distribution of the system. However, such models have to be well defined through a large uv coverage and require a good knowledge of the wind-wind collision zone. We clearly lack this information with our limited data set and we consider this modeling out of the scope of the paper. Nevertheless, we investigate further whether any sign of departure from spherical symmetry of the mid-IR source is seen by comparing the circumstellar dust extension according to the baseline orientation.

\subsection{Departure from spherical symmetry}
Our projected baselines cover position angles from 42$^\circ$ to 105$^\circ$, allowing us to put some constraints on the degree of spherical geometry of the dust. It is also of prime importance to determine as best as possible whether a binary signal is present in the MIDI data and whether the position angles of the projected baselines are aligned with a particular direction reported for the \object{HM\,Sge} binary system from radio, optical or polarization data. 

Let us recall first that even though the WD component has a high luminosity ($\sim$10000L$_{\odot}$) \citep{murs94, murs97} and the binary system separation ($\sim$40mas) is optimum for a detection with the MIDI instrument, its low mid-IR flux prevents MIDI to directly detect its signal ($F^{Mira}_{N band}/F^{WD}_{N band}$$\sim$10$^{5}$). The mid-infrared flux coming from an accretion disk would add to the WD one, but the high temperature of such a structure makes it also too faint in mid-IR. The WD and a putative accretion disk could potentially perturb the dusty environment leading to a large scale signature mimicking a binary signal. Such a signal is probably observed in the case of \object{R\,Aqr} \citep{tuth00}, that the authors attribute to a compact mid-IR source representing 3\% of the N band flux\footnote{Note that this signal is also seen in recent MIDI data.} at 700 mas from the cool star (this separation would scale down to below 100 mas at 1.5kpc). Such a feature might be some localized concentration of dust, a local warming or a global asymmetry of the dust shell as also seen and discussed in Sec.\ref{sect:Vismodels} in the frame of the B01 double shell model that predicts such a binary signature.

\begin{figure}
\includegraphics[width=8.cm]{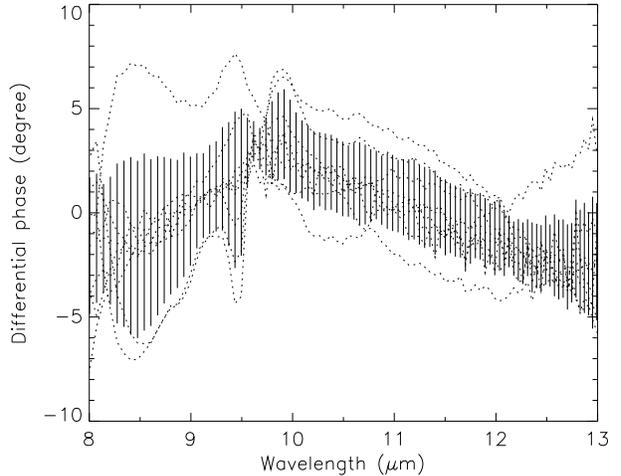}
\caption{\label{diff-phase}Calibrated differential phase of the object evaluated from all the determined projected baselines (dotted lines). Error bars correspond to the rms of the differential phase averaged over all the projected baselines.}

\end{figure}

As previously said, the MIDI visibilities are rather smooth and no hint of such oscillation (as seen for instance in \citet{ches06}) is found in the data (the shape of the visibility profile may uniquely be related to the dust shell). It is possible to be more sensitive to a weak signal of this kind by using the differential phase. Fig.~\ref{diff-phase} shows the calibrated differential phase of the object evaluated from all the determined projected baselines (dotted lines). Error bars correspond to the rms of the differential phase over all the projected baselines. The rms values of the differential phase ($\pm$4.26$^\circ$ at 8.5$\mu$m, $\pm$1.79$^\circ$ at 10.5$\mu$m and $\pm$1.66$^\circ$ at 12.5$\mu$m) indicates that there is no obvious signature of any astrophysical effects. The most discrepant curves are not correlated to a particular baseline length nor position angle and seem to reflect fast atmospheric changes during observations that were not fully corrected by the calibrator. 

In order to estimate further the constraints provided by such measurements, we performed a perturbation of our best single shell model by a point-like structure at varying flux contrast, distance and angle. Our tests allow us to discard any compact mid-IR source representing more than about 1\% of the total flux (i.e. 0.5 to 1 Jy between 8$\mu$m and 13$\mu$m) at a distance ranging from 20mas to 80mas from the Mira star\footnote{With a separation of 40mas, at full sinusoidal pattern should be seen in the spectrally-dispersed MIDI visibilities and differential phases from 8$\mu$m to 13$\mu$m.}.

\begin{figure}
\includegraphics[width=9cm]{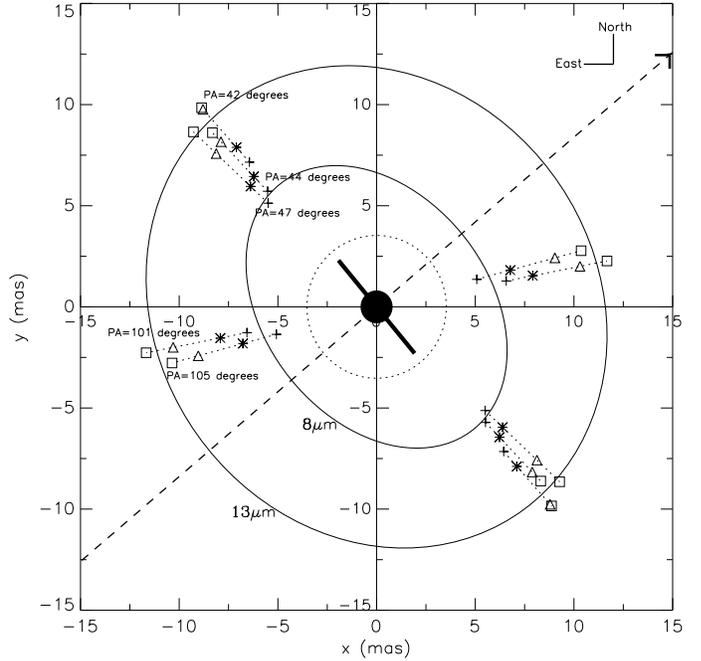}
\caption{Equivalent HWHM of the Gaussian intensity distribution (in mas) dependence on the projected baseline position angle on the sky. These HWHM are taken at 8 (plus), 9 (cross), 10 (triangle) et 13$\mu$m (square). The figure is arbitrarily centered on the Mira and presents the binary axis (dashed line) (PA=130$^\circ$) \citep{eyre01} where the position of the White Dwarf at 40\,mas \citep{eyre01}, represented by the arrow, is out of the window. The thick line, perpendicularly to the binary axis, represents the position angle of the Raman line polarization \citep{schm00}. The black filled circle represents the dimension of the Mira determined with the best perturbed model. The dotted circle represents the dust shell inner boundary dimension determined with the best perturbed model. The two 8 and 13$\mu$m solid ellipses have a minor axis of 5.6 and 11.1 mas, and a major axis of 7.8 and 12.5 mas, respectively.}
\label{equivalent-UD}
\end{figure}

Let us now consider the level of large scale asymmetry for the source. Fig.~\ref{equivalent-UD} shows the equivalent HWHM of the Gaussian intensity distribution according to the orientation of the projected baseline on the sky (UT2-UT3 and UT3-UT4). The figure is centered on the Mira and presents the binary axis direction (around 120-130$^\circ$ according to \citet{eyre01,schm00}) located beyond the window at 40\,mas north-west \citep{eyre01}. The MIDI baselines at PA$\sim$45$^\circ$ are oriented almost perpendicular to binary separation, in the direction of the outflows evidenced by various authors \citep{solf84,hack93,rich99,corr99,eyre01}. On the contrary, the baselines at PA$\sim$100$^\circ$ are favorably oriented to probe the largest extension of the dust, if this dust is elongated in the preferential direction of the shadow of the Mira, that should efficiently protect the dust from the hard emission of the WD. The orbital period of \object{HM\,Sge} is unknown but certainly rather long and we assume that it is barely unchanged since 1998, date of the latest polarization and HST observations\footnote{We notice that the proper motions reported for some extended features in \citet{rich99} is typically 2.5$^\circ$ per year from north to east. If this estimation can be extrapolated to the binary motion, then the rough expected position angle of the projected binary plane should be around 145-150$^\circ$ at the time of the MIDI observations.}. 

Using the Gaussian HWHM estimations, the following ratio between the HWHM at 105$^\circ$ and the ones at 42$^\circ$ can be computed at 8, 9, 10 and 13$\mu$m, respectively: 0.72$^{+0.32}_{-0.19}$, 0.80$^{+0.20}_{-0.17}$, 0.84$^{+0.19}_{-0.15}$ and 0.91$^{+0.14}_{-0.12}$. The extension of the shell is larger in the perpendicular direction than in the binary axis. The level of discrepancy is larger at 8$\mu$m and steadily decreases toward longer wavelengths, with HWHM from 5 to 10\,mas at 8$\mu$m, to about 11 to 13\,mas at 13$\mu$m.

The wind from the cool star is likely to have an equatorial density enhancement. Spectro-polarimetric observations show that continuum from the Mira star is intrinsically polarized \citep{schm00} and this polarization would be, according to the author, most likely due to light scattering in an aspherical photosphere or wind.

\begin{figure}
\includegraphics[width=9cm]{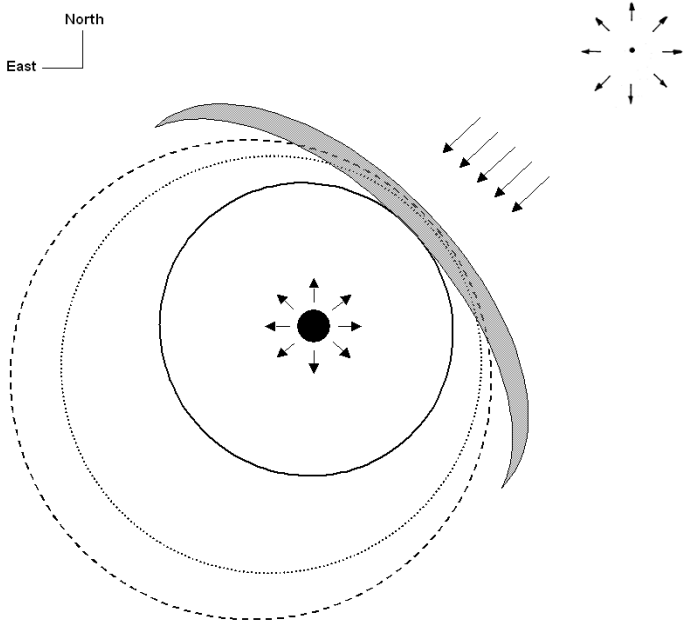}
\caption{Schematic representation of the symbiotic system in which the Mira (in the center of the figure) is wrapped into a shock cone. In the top on the right is the White Dwarf. The arrows surrounding the 2 components indicate the 2 winds and the parallel arrows symbolize the radiation reaching the shock front component (gray filled zone). The contribution of the shock component decreases from 8 to 13$\mu$m simultaneously with the level of asymmetry. The 3 circular regions correspond to the dust shell extension at 8 (solid line), 10 (dotted line), and 13$\mu$m (dashed line).}
\label{geo-cartoon}
\end{figure}

Several geometries can be envisaged:

\begin{itemize}
\item A disk-like geometry seen almost edge-on. We expect to detect a strong difference between the MIDI visibilities with perpendicular baselines that is not observed. Moreover, depending on the inclination and the disk vertical extension, a large circumstellar absorption of the Mira light is expected.
\item A disk-like geometry seen almost pole-one offer an elegant solution to the opacity problem: the light from the Mira star can reach us through an optically thin path whereas a dense and compact disk seen pole-on emits a large infrared excess. This geometry is relatively symmetrical and exhibit similar extensions whatever the direction of the baseline is.
\item A slightly more complex geometry in which the Mira is wrapped into a shock cone (Fig.\ref{geo-cartoon}) as presented in several studies \citet{form95, mast99, kenn05}. The density is not uniform and the envelop may be thinner (and patchy) in the direction perpendicular to the binary direction, allowing the Mira light to reach us, and a large dust content, mostly symmetrical to reside close to the Mira. This geometry is identical for pole-on or edge-on inclinations.
\end{itemize}

The probably high degree of symmetry of the source favors the two last scenario in which most of the dust resides in a spherical/disk-like pole-on envelope around the Mira, that is currently only slightly perturbed by the current hard flux of the WD.
Such a low level of perturbation may appear surprising considering the strong uv source around the cool star, but the separation is rather large ($\sim$60AUs), and the shielding of the uv flux by the dense and slow Mira wind is probably important (see the case of the symbiotic nebula \object{PN M2-9} in \citet{livi01}).

A continuum of inclinations is possible but the observations of a symmetrical optical and radio outflow suggest a high inclination. In the frame of the wind-wind collision geometry, the larger extensions perpendicular to the binary axes could be explained by a somewhat opened aperture angle of the shock cone. At 8$\mu$m, hot dust emits preferentially close to the inner diameter of the Mira shell, but an additional emission component from the shock front may provide the asymmetry observed. As the contribution of the shock component decreases from 8 to 13$\mu$m, the bulk of the emission is more centered around the Mira and the level of asymmetry decreases (see Fig.\ref{geo-cartoon}).

We note that the wind-wind collision hypothesis has been investigated and modeled many times in the literature of \object{HM\,Sge} \citep{eyre01, rich99, gira87, will84a, will84b}. The technique of mid-IR interferometry is a good opportunity to extensively study such a structure, but the present paper shows that two-telescopes observations, as currently performed by MIDI, are not sufficient to constrain its shape efficiently. What is clearly needed is an interferometric imager able to recombine simultaneously the light from the 4 UT or AT telescopes in the mid-IR. Such a recombiner called MATISSE is proposed for the second generation instrumentation of the VLTI \citep{lope06}.

\section{Summary and conclusion}
\label{sect:conclusion}

We have presented new high spatial resolution observations of the dusty environment of \object{HM\,Sge}, consisting on MIDI/VLTI spectra and visibilities. Using the strong constraints provided by optical interferometry, we reassessed previous interpretations based only on the modeling of the spectral density distribution. In the case of the complex symbiotic systems, the spectral energy density can be modeled in different ways. \citet{schi01} and \citet{bogd01} have worked the same year on the ISO/SWS spectra of \object{HM\,Sge}. Both take into account single and double shell models, as an attempt to take into account the spatial complexity of the source. Both used the same radiative transfer code DUSTY. However, they do not converge to the same value of parameters and their respective conclusion  are in complete opposition. The interferometric study of this object shows that the morphology of the dust was close enough to the single shell model of \citet{schi01} considering that the circumstellar environment of the symbiotic system was compact and optically thick (an angular inner radius of the dust shell located at 3.5\,mas and a visual optical depth of 29). 

An improvement of the model consisting in small perturbations of the \citet{schi01} parameters shows that a good fit of both SEDs and visibility curves is obtained steepening the density power law coefficient from 1.8 to 2.2 and decreasing the 10$\mu$m optical depth from 2.47 to 2.15. The change of these two parameters gives a larger transparency of the circumstellar medium and allows the emission to come from deeper layers with the aim to spatially concentrate the emission of the dust shell. Finally the insertion of carbon material in the dusty layers allows to diminish a too much extended structure between 8 and 10$\mu$m. Such a chemical component could come from the White Dwarf ejecta during the outburst of 1975 or from particular dust nucleation processes that occur under the non-equilibrium physical conditions that prevail around the White Dwarf. 

The compact model was rejected by \citet{schi01} because of its high optical depth not compatible with non-absorbed emission of the Mira in the near-IR. Our best spherical model has to face the same objection, but this issue could be solved assuming a more complex spatial distribution of the dust. Our MIDI data allowed us to investigate the departure of spherical symmetry using different baseline orientations. The extension is stronger in the direction perpendicular to the binary axis. This asymmetry is largest at 8$\mu$m and steadily decreases toward longer wavelengths. We speculated on the origin of the departure. We propose that the Mira is wrapped into a shock cone where the density of the envelope is thinner and patchy in the direction perpendicular to the binary axis. This could allow the Mira light to reach us, and a large dust content, mostly symmetrical to reside close to the Mira. Further VLTI/MIDI observations at 15 different projected baselines with ATs are planned to confirm this geometrical scenario.

The MIDI observations must be included in the frame of the long-term evolution of the dusty envelope of \object{HM\,Sge} associated with the variability of the conditions of dust survival in the system, and hence the luminosity and temperature of the White Dwarf, and also the relative position of the Mira and the WD in the orbit. The current dusty envelope from the Mira appears only slightly perturbed by the hot companion. As the time passes, the impact of the WD ionizing flux on the Mira dust shell should decrease further for this large separation binary system unless that in the course of the putatively eccentric orbit a decrease of the separation occurs, or that a large accretion disk builds up.


\begin{acknowledgements}
Walter Jaffe and Rainer K\"ohler are warmly thanked for their constant efforts for improving the reducing software of MIDI, that contributes strongly to the success of the instrument. Without them, this study would not have been possible. Mr. Sacuto benefits from a PHD grant from the Conseil R\'egional Provence - Alpes - C\^ote d'Azur (FRANCE) managed by ADER-PACA.
\end{acknowledgements}

%
%

%
%

\appendix

\section{Determination of the model visibilities}

\begin{figure}
[ptb]
\begin{center}
\includegraphics[width=8.5cm]{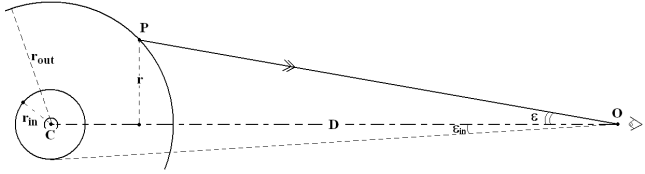}
\caption{Geometry used for the determination of the emergent intensity from the shell outer boundary ($P$) and received under the angle $\varepsilon$ by the observer located at $O$, distant of $D$ from the center $C$ of the central source. The point $r$ is the impact parameter; $r_{in}$ is the inner boundary radius, and $\varepsilon_{in}$ is the angular radius of the inner shell boundary.}
\label{geometry intensity figure}
\end{center}
\end{figure}

The DUSTY code is used to get a theoretical model of the normalized radial intensity distribution at specified wavelengths. For each model, DUSTY provides the radial intensity (in Jy.arcsec$^{-2}$) versus the invariant dimensionless reduced impact parameter $\rho$ defined by

\begin{equation}
\centering
\rho=\frac{r}{r_{in}}
\label{rho parameter}
\end{equation}
where $r$ is the impact parameter and $r_{in}$ is the radius of the inner boundary (see Fig.\ref{geometry intensity figure}).

Taking into account that the star in the code is normalized at luminosity $L_{\star}^{D}=10^{4}L_{{\odot}}$, the radius $r_{in}$ is given by

\begin{equation}
\frac{r_{in}}{r_{in}^{D}}=\sqrt{\frac{L_{\star}}{L_{\star}^{D}}}.
\label{inner radius vs luminosity}
\end{equation}
where $r_{in}^{D}$ is the radius of the inner boundary when $L_{\star}^{D}=10^{4}L_{{\odot}}$ and $L_{\star}$ is the luminosity of the star.

\bigskip Therefore, the impact parameter $r$ is given by

\begin{equation}
r=\rho r_{in}=\rho r_{in}^{D}\sqrt{\frac{L_{\star}}{L_{\star}^{D}}}.
\label{real impact parameter}
\end{equation}
where the luminosity corresponds to the one given by the two authors for the Mira and the White Dwarf (see Tab.\ref{single shell table} and Tab.\ref{two shell table}).

DUSTY also allows us to generate the theoretical visibility curves by calculating the Hankel transform of the circularly symmetric intensity distribution. The code gives the visibility versus the invariant dimensionless scaled spatial frequency $\phi$ defined by

\begin{equation}
\phi=f.2\varepsilon_{in},
\label{scaled spatial frequency}
\end{equation}
where $f$ is the spatial frequency (in arcsec$^{-1}$) and $\varepsilon_{in}$ is the angular radius of the shell inner boundary (see Fig.\ref{geometry intensity figure}).

\bigskip 
Taking into account that DUSTY considers a star with an observed bolometric flux $F_{bol}^{D}=10^{-6}$ W.m$^{-2}$, the angular radius of the inner boundary is given by

\begin{equation}
\frac{\varepsilon_{in}}{\varepsilon_{in}^{D}}=\sqrt{\frac{F_{bol}}{F_{bol}^{D}}},
\label{inner angular radius vs bolometric flux}
\end{equation}
where $\varepsilon_{in}^{D}$ is the angular radius of the inner boundary when $F_{bol}^{D}=10^{-6}$ W.m$^{-2}$ and $F_{bol}$ is the observed bolometric flux of the source.

Therefore, the spatial frequency $f$ is given by

\begin{equation}
f=\frac{\phi}{2\varepsilon_{in}}=\frac{\phi}{2\varepsilon_{in}^{D}}\sqrt{\frac{F_{bol}^{D}}{F_{bol}}}.
\label{real spatial frequency}
\end{equation}

DUSTY can produce maps of the visibility at up to 20 wavelengths. The visibilities of 20 spectral channels are generated between 7.5 and 13$\mu$m corresponding approximately to the number of independent channel provided by MIDI with the low resolution mode (prism, R$\sim$20).

\section{The binary separation}
\begin{figure}
\centering
\includegraphics[width=8cm]{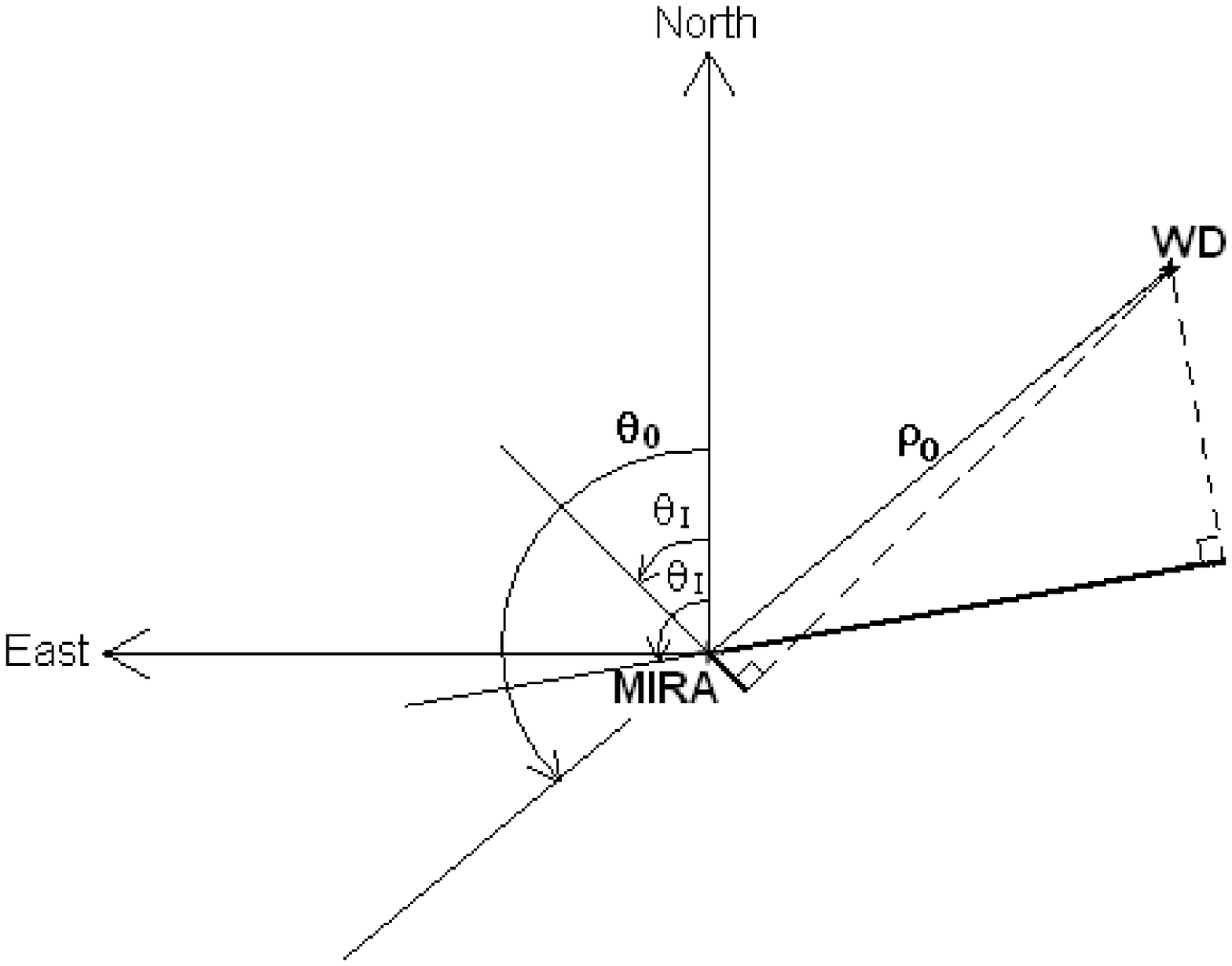}
\caption{Schematic view of the separation between the binary components. $\rho_{0}$=40\,mas is the distance between the Mira and the White Dwarf (WD) and $\theta_{0}$=130$^\circ$ is the position angle of the binary axis \citep{eyre01}. $\theta_{I}$ are the position angles (at 45$^\circ$ and 100$^\circ$ from the north to the east) of the interferometric baselines from the sky. The thick lines correspond to the projected separations of the binary components on the sky.}
\label{bina_sepa}
\end{figure}

In the frame of a two shell model, the relative position of the centers has to be taken into account in the computation of the visibilities.
We adopted a binary separation and a position angle of the binary axis estimated of 40\,mas and 130$^\circ$ respectively \citep{eyre01}.
Putting the Mira star at the center of reference (see Fig.\ref{bina_sepa}), the expression of the visibility can be written as:

\begin{eqnarray}
V=\frac{\sqrt{C_{tot}C^{\star}_{tot}}}{F_{tot}}= & \\ \nonumber
\frac{\sqrt{C^{2}_{M}+C^{2}_{WD}+2C_{M}C_{WD}cos(2\pi\frac{B_{I}}{\lambda}\rho_{0}cos(\theta_{I}-\theta_{0}))}}{F_{M}+F_{WD}}
\label{visibility}
\end{eqnarray}
where $C_{tot}$ is the total correlated flux of the Mira plus the White Dwarf ($C_{tot}=C_{M}+C_{WD}e^{-2\pi i\frac{B_{I}}{\lambda}\rho_{0}cos(\theta_{I}-\theta_{0})}$) and $C^{\star}_{tot}$ is its conjugate.
$\overrightarrow{B}$=($B_{I}$;$\theta_{I}$) are the coordinates associated to the interferometer projected baseline on the sky where $B_{I}$ is the baselength and $\theta_{I}$ its position angle from the north to the east. $\overrightarrow{\rho}$=($\rho_{0}$;$\theta_{0}$) are the coordinates associated to the binary components where $\rho_{0}$ is the binary separation and $\theta_{0}$ the position angle of the binary axis from the north to the east. Finally, $F_{tot}$ represents the total flux of the Mira plus the White Dwarf ($F_{tot}=F_{M}+F_{WD}$).

\end{document}